\documentclass[11pt]{article}
\usepackage{graphicx,mathtools,amssymb,bbm,breqn,amsxtra,amsthm,float,authblk}
\usepackage{subcaption}
\usepackage[ruled,vlined]{algorithm2e}
\usepackage{algorithmic}

\usepackage[textsize=tiny]{todonotes}

\usepackage[margin=1.0in]{geometry}
\usepackage{sectsty}
\usepackage{hyperref}
\newtheorem{theorem}{Theorem}

\newcommand{\R}{\mathbb{R}}
\newcommand{\Ex}{\mathbb{E}}
\newcommand{\Pb}{\mathbb{P}}
\newcommand{\de}{\text{d}}

\newcommand{\ddt}{\frac{\partial}{\partial t}}
\newcommand{\Var}{\mathbb{V}}
\newcommand{\A}{\mathbf{A}} 
\newcommand{\B}{\mathbf{B}}
\newcommand{\He}{\text{He}}

\newcommand{\bfA}{\textbf{A}}

\newcommand{\Id}{\textbf{I}}

\newcommand{\revi}[1]{\textcolor{black}{#1}}

\graphicspath{{Figures/}}

\usepackage{titlesec}
\titleformat*{\section}{\normalsize\bfseries}
\titleformat*{\subsection}{\normalsize\bfseries}
\titleformat*{\subsubsection}{\normalsize\bfseries}
\titleformat*{\paragraph}{\normalsize\bfseries}
\titleformat*{\subparagraph}{\normalsize\bfseries}

\begin{document}

\title{\Large{\textbf{A Koopman framework for rare event simulation in stochastic differential equations}}}
\author[1]{Benjamin J.~Zhang\footnote{Corresponding author. Email: \texttt{bjz@mit.edu}}}
\author[2]{Tuhin Sahai}
\author[1]{Youssef M.~Marzouk}
\affil[1]{\small{Department of Aeronautics \& Astronautics}\\ \small{Center for Computational Science \& Engineering}\\ \small{Massachusetts Institute of Technology} }
\affil[2]{\small{Raytheon Technologies Research Center}}

\date{31 January 2022}
\maketitle

\abstract{
We exploit the relationship between the stochastic Koopman operator and the Kolmogorov backward equation to construct importance sampling schemes for stochastic differential equations. Specifically, we propose using eigenfunctions of the stochastic Koopman operator to approximate the Doob transform for an observable of interest (e.g., associated with a rare event) which in turn yields an approximation of the corresponding zero-variance importance sampling estimator. Our approach is broadly applicable and systematic, treating non-normal systems, non-gradient systems, and systems with oscillatory dynamics or rank-deficient noise in a common framework. In nonlinear settings where the stochastic Koopman eigenfunctions cannot be derived analytically, we use dynamic mode decomposition (DMD) methods to \revi{approximate} them numerically, but the framework is agnostic to the particular numerical method employed. Numerical experiments demonstrate that even coarse approximations of a few eigenfunctions, where the latter are built from non-rare trajectories, can produce effective importance sampling schemes for rare events. 
}

\section{Introduction}

Understanding and quantitatively characterizing rare phenomena is important to modeling, design, and decision making in a variety of science and engineering disciplines. Examples include studying the failure of materials \cite{liu2015efficient}, predicting the insolvency of financial institutions \cite{embrechts1982estimates}, understanding the occurrence of rogue waves \cite{dematteis2018rogue,cousins2019predicting}, estimating reaction rates in computational chemistry \cite{vanden2006transition}, and assessing the reliability of aerospace systems \cite{zhang2018rare}. Many of these examples involve dynamical systems forced by random noise, which is captured in the form of Brownian motion, and a key challenge is to compute the probabilities of noise-induced rare events and the predominant mechanisms by which they occur. These rare events are often associated with adverse outcomes and failures.

In this paper, we present a general framework for constructing Monte Carlo estimators of rare event probabilities, and of other expectations associated with rare events, in nonlinear stochastic differential equations (SDEs). Rare event simulation for SDEs is particularly challenging for two reasons. It first requires a faithful model, i.e., one that exhibits the rare phenomena of interest with sufficiently accurate probability. Second, a computationally efficient methodology is needed to produce the rare event, i.e., to characterize the tails of the relevant distributions. Performing the latter also elucidates the pathways or \emph{mechanisms} leading to a rare event. 

For SDEs, expectations with respect to the induced path-space probability measures are, generally, difficult to compute directly. Hence Monte Carlo methods are often used to estimate these expectations instead. Simple Monte Carlo methods, while robust, are inefficient for estimating expectations sensitive to rare events \cite{vanden2012rare}. Since rare events by definition occur infrequently, the variance of a simple Monte Carlo estimator can be very large relative to the quantity of interest. Furthermore, for rare events in SDEs that obey a large deviations principle, simple Monte Carlo methods require an exponentially increasing number of samples to maintain a constant relative error as the noise factor in the Brownian motion decreases linearly \cite{dembo1998large,vanden2012rare}. For these reasons, a vast body of literature has focused on devising sampling methods that improve on simple Monte Carlo for rare event simulation \cite{budhiraja2019analysis}.

Importance sampling for SDEs constitutes a major class of Monte Carlo methods for simulating rare events. Here, one simulates an alternative dynamical system whose trajectories reach the rare event more often. Each of these samples is then re-weighted according to its importance relative to the original SDE's distribution. These weights are given by the celebrated Girsanov theorem \cite{oksendal2003stochastic}. Multi-level splitting and subset simulation comprise a different class of adaptive Monte Carlo methods for rare events in which, over a series of iterations, one creates an artificial drift towards the rare event of interest. Splitting methods were first conceived in \cite{kahn1951estimation}, with more computationally efficient methods proposed recently \cite{villen1991restart}. Subset simulation, on the other hand, was originally proposed in the engineering reliability literature \cite{au2001estimation} and has been widely adopted and improved upon by the civil engineering community \cite{papaioannou2015mcmc}. While subset simulation can be used to estimate rare event probabilities in dynamical systems, it is typically used to sample static models. Links between subset simulation and sequential Monte Carlo are described in \cite{cerou2012sequential}, and multilevel splitting has been extended to static and non-Markovian models in \cite{botev2012efficient}. Importance sampling for SDEs, while simple to implement and easily parallelizable, is intrusive: in the context of SDEs, it requires altering the drift term of the model, which may be impossible when the model is given as a black box. In contrast, multilevel splitting methods are applicable in black-box settings and often more stable than their importance sampling counterparts \cite{budhiraja2019analysis}. That said, splitting is generally more difficult to implement than importance sampling. Furthermore, to the authors' knowledge, there is no simple characterization of a multi-level splitting scheme that yields a zero-variance estimator. 

Importance sampling and multilevel splitting have been further enhanced by large deviations theory \cite{dembo1998large,varadhan1984large}. These methods exploit the large deviations principle as an alternative mechanism of characterizing rare events in dynamical systems, inform the implementation of splitting and importance sampling, and provide theoretical guarantees on estimator efficiency \cite{vanden2012rare,dupuis2004importance,dean2009splitting}. These methods also appear in the literature as so-called \emph{instanton}-based sampling methods, where minimizers of the system's large deviations rate function describe how to push the system towards the rare event of interest \cite{ebener2019instanton,margazoglou2019hybrid}. This approach was first noted in queuing theory \cite{siegmund1976importance}, but has also been used for problems in the physical sciences. These enhancements are also related to \emph{variational} approaches to importance sampling, where the alternative SDE is posed as the solution to a stochastic optimal control problem---which, in principle, can yield zero-variance estimators \cite{hartmann2017variational,hartmann2019variational,kebiri2018adaptive,zhang2014applications}. The drawbacks of these approaches are also well-noted. Large deviations-based approaches for sampling are optimal in an asymptotic sense, but counter-examples have been constructed to show that they can lead to larger variance than applying direct Monte Carlo \cite{glasserman1997counterexamples}. And the computational effort required to solve stochastic optimal control problems can be untenable in high-dimensional settings.

In this paper, we propose a novel framework for rare event simulation that uses tools originating from dynamical systems theory and combines them with importance sampling. Specifically, we show that the \emph{stochastic Koopman eigenfunctions} associated with a given SDE can be used to {accurately} and {efficiently} approximate zero-variance importance sampling (IS) estimators. The approach leverages recent developments in Koopman operator approximation techniques and only assumes that the SDE is amenable to numerical Koopman analysis. We emphasize that our framework is agnostic to the numerical method used for approximating the Koopman operator. 

The last decade has witnessed considerable interest in operator-theoretic and data-driven computational approaches for analyzing and manipulating nonlinear dynamical systems. The Koopman operator is a linear mapping on the space of observables of a given dynamical system \cite{koopman1931hamiltonian,budivsic2012applied,mauroy2020koopman}. Its existence provides a \emph{global} linearization of the dynamics and enables spectral analysis for nonlinear systems. Moreover, the discovery that data-driven methods for dynamical systems such as dynamic mode decomposition (DMD) \cite{schmid2010dynamic} (originally conceived in the fluid mechanics community) can effectively \revi{approximate} spectral objects of the Koopman operator \cite{rowley2009spectral} has led to their further development and widespread application. 

In this work, we exploit the relationship between zero-variance sampling and the Koopman operator to show that DMD methods can be integrated with importance sampling to create new rare event simulation techniques. Our framework provides a systematic approach with \emph{general} applicability. For example, existing rare event simulation techniques are often demonstrated on gradient systems, or on systems with normal dynamics. Our approach is also applicable to non-gradient systems, non-normal systems that display transient growth, and oscillatory dynamics. This flexibility is critical for extending efficient dynamic rare event simulation to realistic engineering problems \cite{zhang2018rare}.

A key feature of our approach is that we leverage the data-driven nature of Koopman numerics to provide insight into rare events via simulation of \emph{non-rare trajectories}. The ability to resolve Koopman eigenfunctions near the rare event using non-rare trajectories enables computation of a biasing that ``pushes'' importance sampling trajectories into the rare event regions. We show that even \emph{coarse approximations} of the Koopman eigenfunctions using non-rare trajectories can produce good importance sampling estimators for rare event simulation. The method is asymptotically exact in the sense that as one employs a larger number of Koopman eigenfunctions, the variance of the corresponding importance sampling estimator tends towards zero. We provide a non-asymptotic analysis that describes how, under certain conditions, the second moment of the importance sampling estimator is bounded by a term that depends on how well the Koopman eigenfunctions approximate the observable of interest. 

\subsection{Problem setting and notation}

Let $\{X_t\}_{t\in[0,T]}$ be a time-homogeneous diffusion process evolving according to the SDE,
\begin{align}
	\begin{cases}
		\de X_t &= \textbf{A}(X_t )\, \de t +\textbf{B}(X_t )\, \de W_t \\
		X_0 &= x,
	\end{cases}
	\label{eq:sde}
\end{align}
where $X_t$ is an element of $\R^d$, $\textbf{A}$ is a function from $\R^d$ to itself, $\textbf{B}$ is a function from $\R^d$ to the space of $d\times r$ real-valued matrices, and $W_t$ is a standard $r$-dimensional Brownian motion. To guarantee existence and uniqueness of a strong solution to the SDE, we assume the drift vector and diffusion matrix are locally Lipschitz in space \cite{karatzas2012brownian}. We wish to estimate
\begin{align}
	\rho=  \Ex\left[f(X_T)|X_0 = x \right] = \int_{\R^d} f(x) \pi_T(x) \,\de x,
	\label{eq:qoi}
\end{align}
where $f(x)$ is non-negative, 
and $\pi_T(x)$ is the probability density of the state at time $T$. Note that if $f(x)$ were an \emph{indicator function over some rare event of interest} $E$, then $\rho$ would equal the probability of the state being in region $E$ at time $T$. We also assume that the system has an invariant distribution $\eta_\infty$.

In the next section, we review some theoretical tools for importance sampling in stochastic differential equations. In Section~\ref{sec:iskoopman}, we discuss the Koopman operator and related numerical methods, and we present our framework for constructing importance sampling estimators. In Section~\ref{sec:numericalexamples}, we demonstrate the methodology on a range of illustrative stochastic dynamical systems. We analyze the variance of the importance sampling estimators produced by our methodology in Section~\ref{sec:analysis}.  We conclude and discuss future work in Section~\ref{sec:discussion}.

\section{Rare event simulation for SDEs}

We start with an overview of analytical tools for studying stochastic differential equations, including the infinitesimal generator and the Kolmogorov equations. Much of this discussion is based on Karatzas and Shreve \cite{karatzas2012brownian}, {\O}ksendal \cite{oksendal2003stochastic}, and Pavliotis \cite{pavliotis2014stochastic}. We also review importance sampling in the context of SDEs and describe related approaches to rare event simulation based on stochastic optimal control and large deviations theory. 

\subsection{Kolmogorov equations}
Let $X_t$ be defined by the SDE in \eqref{eq:sde}. One of the primary tools for studying stochastic processes is the infinitesimal generator defined as
\begin{align}
	\mathcal{A}f = \lim_{t\to 0} \frac{\Ex \left[f(X_t) \vert X_0 = x \right]-f(x)}{t},
	\label{eq:generator}
\end{align} 
for $f\in\mathcal{D}_{\mathcal{A}}$, where $\mathcal{D}_{\mathcal{A}}$ is the set of functions for which the above limit exists for all $x\in \R^d$. For SDEs, a closed form expression of the limit involves the drift and diffusion terms as follows,
\begin{align}
	\mathcal{A}f &= \langle \textbf{A}(x),\nabla f \rangle + \text{Tr}\left[\textbf{Q}(x) \nabla^2 f \right],	\label{eq:sdegen}
 \\
	& = \sum_{i = 1}^d A_i(x)\frac{\partial f}{\partial x_i} + \sum_{i = 1}^d \sum_{j = 1}^d  Q_{ij}(x) \frac{\partial^2 f}{\partial x_i \partial x_j}, \nonumber
\end{align}
where $\textbf{Q}(x) = \frac{1}{2}\textbf{B}(x)\textbf{B}(x)^*$ and $f$ is a twice-continuously differentiable function on $\R^d$. The infinitesimal generator appears in the Kolmogorov equations, which are two PDEs that describe the evolution of densities and statistics of a given SDE. The \emph{Kolmogorov backward equation} (KBE) describes the time-evolution of expectations of functions of the state. Let $\Phi(t,x) = \Ex^{t,x}\left[ f(X_T)\right]\coloneqq\Ex\left[ f(X_T) | X_t = x\right]$ be defined on $t\in[0,T]$, where $T>0$. Then
\begin{align}
\begin{dcases}
	\frac{\partial \Phi}{\partial t} + \mathcal{A}\Phi  = 0 \\
		\Phi(T,x) = f(x).
\end{dcases}
\label{eq:kbe}
\end{align}
The \emph{Kolmogorov forward equation} (KFE), also known as the Fokker--Planck equation, describes the evolution of the probability density function of the state. The equation is found by considering the $L^2$-adjoint of the infinitesimal generator. Let $\pi(t,x)$ be the probability density of $X_t$. Then
\begin{align}
\begin{dcases}
	\frac{\partial \pi}{\partial t} = \mathcal{A}^* \pi(t,x) \\
	\pi(0,x) = \pi_0(x) 
\end{dcases}
\end{align}
where the adjoint is
\begin{align}
	\mathcal{A}^* \pi = -\nabla \cdot\left(\textbf{A}(x) \pi \right) + \text{Tr}\left[\nabla^2(\textbf{Q}(x) \pi) \right].
\end{align}
Theoretically, expectations such as \eqref{eq:qoi} can be found via a direct solution of the KBE. The quantity of interest is simply an evaluation of the solution: $\rho = \Phi(0,x)$. However, solving the KBE exactly is expensive and increasingly intractable as the dimension of the state space grows. Furthermore, when one is interested in quantities such as rare event probabilities, the required solution accuracy typically becomes prohibitive. For this reason, we turn to sampling methods, in which multiple independent simulations of an SDE are performed to estimate expectations through a sample average. While a direct solution of the Kolmogorov equations may not be feasible, in what follows, we show that these equations can be used to approximate zero-variance estimators. 

\subsection{Importance sampling for SDEs}
We now review some basic notions of Monte Carlo and importance sampling methods for SDEs. Let $\mathbb{P}$ be the path-space measure induced by the SDE in \eqref{eq:sde}. A simple Monte Carlo method for estimating $\rho$ involves generating $M$ independent simulations of the SDE, evaluating the function of interest $f(x)$ at the end of each sample path, and computing the sample average. We then have 
\begin{align}
	\rho \approx \hat{\rho} = \frac{1}{M}\sum_{i = 1}^M f(X_T^{(i)}),
\end{align}
where the samples $X^{(i)}$ are drawn independently from $\mathbb{P}$. The efficiency of a Monte Carlo estimator is typically evaluated by considering its variance and relative error (also known as the coefficient of variation, i.e., the standard error divided by the quantity of interest) \cite{asmussen2007stochastic}. They are, respectively,
\begin{align}
	\mathbb{V}\left[\hat{\rho} \right] &= \frac{1}{M} \Var \left[ f(X_T) \right],  \\
	r_e &= \frac{1}{\rho}\sqrt{\mathbb{V}\left[\hat{\rho} \right]}.
	\label{eq:varrelerr}
\end{align} 
 A good unbiased estimator should have low variance, but when estimating a small $\rho$ such as a rare event probability, relative error is the better metric. This is because $r_e$ can still be large if $\rho$ is orders of magnitude smaller than $\Var[\hat{\rho}]$. In other words, our goal is to ensure that the standard deviation of the estimator scales in proportion with the probability of interest. \revi{The relative error per sample, $r_e \sqrt{M}$, is a useful standardized measure of performance as it is independent of the sample size $M$ \cite{salins2016rare}.} 

The inefficiency of simple Monte Carlo methods is clear when used to estimate rare event probabilities. Let $f(x) = \mathbbm{1}_E(x)$ where $E\subset\R^d$ is a region of phase (or observable) space visited infrequently. The variance and relative error of the estimator are, 
\begin{align*}
	\Var\left[ \hat{\rho} \right] &= \frac{\rho-\rho^2}{M} \approx \frac{\rho}{M}, \\
	 r_e &\approx\frac{1}{\sqrt{M\rho}}.
\end{align*}
 We can see that the number of samples required to keep the relative error below 1 is $O(1/\rho)$. This task is particularly intractable when it is computationally expensive to procure samples from the dynamical system. In simple Monte Carlo, the only way one can reduce the variance of the estimator is by increasing the number of samples in each estimate. Variance reduction methods pursue different mechanisms for reducing the variance beyond simply increasing the number of samples.

One common variance reduction approach is \emph{importance sampling}, which involves drawing samples from an alternative probability measure, $\mathbb{Q}$, that is absolutely continuous with respect the original probability distribution, such that the variance of the resulting estimator is reduced. To account for the bias introduced when sampling from the alternative probability distribution, each sample is weighted according its relative importance with respect to the original measure $\mathbb{P}$. In particular,
\begin{align}
 \hat{\rho}_{IS} = \frac{1}{M} \sum_{i = 1}^M f(\tilde{X}_T^{(i)}) \frac{\de \mathbb{P}}{\de \mathbb{Q}}(\tilde{X}^{(i)}),
 \label{eq:isest}
 \end{align} 
 where $\tilde{X}^{(i)}$ are drawn independently from $\mathbb{Q}$. The variance of this estimator is dependent on the product of the function $f(x)$ and the likelihood ratio between $\mathbb{P}$ and $\mathbb{Q}$, and on the number of samples drawn:
 \begin{align}
 	\Var[\hat{\rho}_{\text{IS}}] = \frac{1}{M}\Var_{\mathbb{Q}} \left[f(\tilde{X}_T) \frac{\de \mathbb{P}}{\de \mathbb{Q}} \right].
 	\label{eq:variance}
 \end{align}
Thus, designing a measure $\mathbb{Q}$ provides an additional mechanism to reduce the variance of the sampling method. For SDE systems, the only admissible choice of $\mathbb{Q}$ is induced by another SDE system $\{\tilde{X}\}_{t\in[0,T]}$ with the same diffusion term as the original SDE and a different drift term \cite{oksendal2003stochastic,stroock2018elements}:
 \begin{align}
 	\begin{cases}
 		\de\tilde{X}_t = \left[\textbf{A}(\tilde{X}_t) + \textbf{B}(\tilde{X}_t) {u}(t,\tilde{X}_t) \right] \de t + \textbf{B}(\tilde{X}_t) \de W_t\\
 		\tilde{X}_0 = x.
 	\end{cases}
 	\label{eq:sdealt}
 \end{align}
 Here $u(t,x)$ is called the biasing function. This function serves as a feedback controller that guides the system such that the resulting importance sampling estimator has lower variance. The likelihood ratio is now given by Girsanov's theorem \cite{oksendal2003stochastic,karatzas2012brownian}:
 \begin{align}
 	Z(\tilde{X})\equiv \frac{\de\mathbb{P}}{\de\mathbb{Q}}(\tilde{X}) = \exp\left(-\int_0^T \langle {u}(t,\tilde{X}_t), \de W_t \rangle - \frac{1}{2}\int_0^T \| {u}(t,\tilde{X}_t)\|^2 \de t \right).
 	\label{eq:girsanov}
 \end{align}
 
 The task now is to choose ${u}(t,x)$ such that the variance of the resulting importance sampling estimator is smaller, or better yet, zero. Assuming that $f(x)$ is twice-continuously differentiable and strictly positive, there exists a choice of ${u}(t,x)$ that leads to a \emph{zero-variance} importance sampling estimator. This choice is the celebrated \emph{Doob h-transform} \cite{vanden2012rare,rogers2000diffusions,sarkka2019applied}. 

 \begin{theorem}[Doob $h$-transform]
 \label{theo:doob}
 	Let $f \in \mathcal{C}^2$ be strictly positive. Let $\Phi(t,x) = \Ex^{t,x}[f(X_T)]$ be the solution to
 \begin{align}
 \begin{dcases}
 	\frac{\partial \Phi}{\partial t}+ \mathcal{A} \Phi = 0, \\
 	\Phi(T,x) = f(x).
 \end{dcases}
 \end{align}
 	 Then using the biasing function 
 	\begin{align}
 		u(t,x) = \textbf{B}^*(x) \nabla \log \Phi(t,x)
 	\end{align}
 	in \eqref{eq:sdealt} will satisfy 
 	\begin{align}
 		f(\tilde{X}_T) \exp\left[-\int_0^T  \langle u(t,\tilde{X}_t), \de W_t \rangle  - \frac{1}{2} \int_0^T \|u(t,\tilde{X}_t)\|^2 \de t \right] = \Phi(0,x).
 		\label{eq:zerovarest}
 	\end{align}
 \end{theorem} 
The proof is provided in Appendix~\ref{app:doobh}.\footnote{As written, Theorem~\ref{theo:doob} does not apply when $f$ is an indicator function. This is an artifact of the simple way we have chosen to express the result. It is straightforward to modify it by conditioning on the event that $X_T$ enters a particular region; indeed, the Doob transform was originally derived just for conditioned processes \cite{rogers2000diffusions}.
Also, our numerical experiments will mollify \revi{and positivize} the indicator function in constructing numerical approximations of the Doob transform, such that Theorem~\ref{theo:doob} applies directly.}
\revi{For more background on the Doob $h$-transform, we refer the reader to, e.g., \cite{rogers2000diffusions,sarkka2019applied}.} 
This choice of biasing results in a zero-variance estimator for $\rho$ since $\rho = \Phi(0,x)$. This result should not be surprising: having access to the exact solution to the KBE enables construction of a Monte Carlo estimator with zero variance, since an evaluation of the solution is, itself, a zero-variance estimator. Though this relationship might seem tautological, it provides useful insights for devising efficient rare event simulation techniques.

Previous approaches recast this problem in terms of optimal control. By defining a new function $U(t,x) = -\log \Phi(t,x)$, one can obtain a PDE for $U(t,x)$ by performing a change of variables on the KBE. The resulting PDE is known as a \emph{stochastic Hamilton--Jacobi--Bellman} (HJB) equation, which can be reformulated as a stochastic optimal control problem. In \cite{hartmann2017variational}, the authors opt to solve the stochastic optimal control problem directly by using this formulation in conjunction with the Donsker--Varadhan variational formula. One can further recast the problem in terms of the solution of a system of forward-backward SDEs \cite{kebiri2018adaptive}. This approach also admits a cross-entropy interpretation for importance sampling for SDEs \cite{zhang2014applications}. 

A similar approach incorporates the theory of large deviations, specifically the Freidlin--Wentzell theory for small noise diffusions \cite{freidlin1998random}. Here, one considers a noise parameter $\epsilon$ that scales the diffusion term, by replacing $\mathbf{B}(x)$ with $\sqrt{\epsilon}\mathbf{B}(x)$. Then by considering the variable transformation $U^\epsilon(t,x) = -\epsilon\log \Phi(t,x)$ and sending $\epsilon$ to zero, one obtains a Hamilton--Jacobi equation whose solution is related to the large deviations rate function of the system \cite{vanden2012rare}. It was found that \emph{subsolutions} of this Hamilton--Jacobi equation \cite{dupuis2004importance,dupuis2007subsolutions,dupuis2015escaping}, for diffusion processes on $\R^d$ and in function space \cite{salins2016rare}, result in provably asymptotically efficient estimators. However, the drawback of this approach is that the subsolution of the Hamilton-Jacobi equation must be ``guessed,'' which is not always straightforward. Note that exact solutions of the resulting deterministic optimal control problem lead to strongly efficient estimators for SDEs \cite{vanden2012rare}.
 
Our approach, described below in Section~\ref{sec:iskoopman}, will avoid both of the above reformulations by directly computing \emph{approximate Doob transforms} using approximate solutions to the KBE. These solutions of the KBE will be expressed in terms of the eigenfunctions of the stochastic Koopman operator. Our approach can also be related to the work of \cite{chiavazzo2017intrinsic}, in which the authors combine trajectory data from molecular dynamics simulations with nonlinear manifold learning techniques to inform the exploration of rare regions of state space. However, their technique is restricted to gradient systems for computational chemistry applications. 

\subsection{Related rare event problems} 
The problem posed in \eqref{eq:qoi} is just one of many scenarios that are of interest in rare event simulation. In this paper, we only consider the problem of the state being in some region of interest at some fixed future time $T$. Another common problem is to compute the probability of entering some region, $E$, before another, $F$: hence $\Pb(X_\tau \in E)$, where $\tau = \inf\left\{t>0 : X_t \in E \cup F  \right\}$. A variation of this problem considers path-dependent quantities, which involve functionals of sample trajectories. These problems are well-studied in the computational chemistry community, where one seeks rare paths between long-lived molecular configurations \cite{vanden2006transition,hartmann2019variational}. This quantity of interest is associated with the solution of a boundary value problem, and its approximation can also be used for sampling. The application of data-driven dynamical systems methodologies to this problem has been studied in \cite{thiede2019galerkin}. 

Another quantity of interest is the probability of entering some set of interest within a \emph{fixed} finite time interval, i.e., $\Pb(\tau \le T)$ where $\tau = \inf \left\{ t>0 : X_t \in E \right\}$. This problem is associated with escaping from attracting sets of a dynamical system and is well-studied in \cite{dupuis2015escaping,spiliopoulos2015nonasymptotic}. In this case, asymptotically efficient importance sampling estimators are designed by considering an initial-boundary value problem associated with the KBE. 

\section{Importance sampling using the Koopman operator}
\label{sec:iskoopman}

We first review the deterministic and stochastic Koopman operators, and discuss how they can be used to approximate expectations and probabilities. We then describe how we will use the stochastic Koopman operator to construct importance sampling schemes for SDEs.

\subsection{The Koopman operator and its generator}
A traditional approach to analyzing dynamical systems involves simulating the evolution of \emph{states}. The Koopman operator \cite{koopman1931hamiltonian} provides an alternative perspective: it represents the dynamical system in terms of the evolution of \emph{observables}. The key advantage is that the evolution of observables is linear even when the underlying system is nonlinear, thus enabling spectral analysis of nonlinear systems \cite{budivsic2012applied}. 

Let $x_t$ be an autonomous dynamical system on $\R^d$ evolving according to $\dot{x} = a(x)$. Let $F^t$ be the flow map; that is, if $x_0$ is the initial condition, then $x_t = F^t x_0$. Let $f: \R^d \longrightarrow \R$ be an observable in some space of functions $\mathcal{H}$. The Koopman operator (KO) is defined as
\begin{align}
	\mathcal{K}^t f(x) = (f \circ F^t) (x). 
\end{align}
It is trivial to show that the KO is linear even when the dynamical system is nonlinear. This property allows one to study the eigenfunctions and eigenvalues of the operator. A function $\phi(x)$ is a Koopman eigenfunction if it satisfies $\mathcal{K}^t \phi(x) = e^{\lambda t} \phi(x)$, where $\lambda$ is the corresponding Koopman eigenvalue. \revi{In stochastic calculus, the stochastic Koopman operator is known as the Markov semigroup operator of the SDE \cite{pavliotis2014stochastic}.}

The stochastic Koopman operator (sKO) is defined in a similar fashion \cite{vcrnjaric2019koopman}. We focus our attention on random dynamical systems that evolve according to SDEs as defined in \eqref{eq:sde}. Let $\{X_t\}_{t\in[0,T]}$ be a stochastic process and $f$ be a twice continuously differentiable real-valued observable, respectively. Then the stochastic Koopman operator is defined as,
\begin{align}
	\mathcal{K}^t f(x) = \Ex[ f(X_t) | X_0 = x ] = \Ex^{0,x}[ f(X_t) ],
\end{align}
where the expectation is taken over the distribution of the state of the stochastic process at time $t$. Analogous to the deterministic setting, the sKO is also linear, leading to the spectral analysis of nonlinear SDEs. The evolution of the expectation of the sKO's eigenfunctions at future times is simple to determine. If $\phi(x)$ is an eigenfunction of the sKO, then $\Ex[\phi(X_t)|X_0 = x] = e^{\lambda t} \phi(x)$. Thus, the time evolution of certain observables of the dynamical system can be determined computationally.

\subsection{Approximating expectations and probabilities}
Assuming that the sKO eigenfunctions exist and form a basis for a suitable function space, expectations and probabilities associated with an SDE can, in principle, be calculated from all the eigenfunctions. Specifically, we can write the expectation of an observable at some fixed time in terms of the expectations of the sKO eigenfunctions, by first expressing the observable as a \emph{linear combination} of these eigenfunctions.\footnote{For SDEs that admit an invariant measure and whose generators are compact and self-adjoint, the spectral theorem guarantees the existence of eigenvalues and a complete orthonormal set in $L^2(\eta_\infty)$, where $\eta_\infty$ is the invariant measure. A frequently studied class of systems that admits a complete set of eigenfunctions are reversible diffusions. One example of a reversible diffusion occurs when the drift term is the gradient of a potential function and the diffusion matrix is the identity. In these cases, the solutions to the Kolmogorov equations can be found via eigenfunction expansions. See \cite{pavliotis2014stochastic} for further details. Irreversible OU processes with invariant measure $\nu$ have also been shown to admit a complete basis of eigenfunctions on $L^p(\nu)$ for $p > 1$ \cite{metafune2002spectrum}.} 

A finite collection of eigenfunctions can thus provide an approximation to the expectations and probabilities of interest. Let $f$ represent some observable of interest and $\{\phi_i(x)\}_{i = 1}^N$ be a collection of $N$ eigenfunctions of the sKO with corresponding eigenvalues $\{\lambda_i\}_{i = 1}^N$. Approximating the observable in terms of the eigenfunctions gives
\begin{align}
	f(x) \approx \sum_{i = 1}^N f_i \phi_i(x), \label{eq:approxobservable}
\end{align}
and hence, 
\begin{align}
	 \Ex\left[f(X_t) |X_0 = x \right] &\approx \sum_{i = 1}^N f_i \Ex^{0,x}[\phi_i(X_t)],  \label{eq:qoikoopapprox}  \\	
	 &= \sum_{i = 1}^N f_i \mathcal{K}^t \phi_i(x) \nonumber \\
	& = \sum_{i = 1}^N f_i e^{\lambda_i t} \phi_i(x).  \nonumber
\end{align}
For rare event probabilities, it suffices to replace $f$ with an indicator function over the rare set of interest; that is, to compute $\Pb(X_T\in E|X_0 = x)$, one would choose $f(x) = \mathbbm{1}_E(x)$. In general, making this approximation accurate may require \emph{accurately} computing \emph{many} sKO eigenfunctions, which may not be practical in most settings. Instead we can combine this idea with importance sampling, as follows.

The sKO eigenfunctions can be used to create approximate solutions to the Kolmogorov backward equation. For continuous-time autonomous dynamical systems, the set of stochastic Koopman operators $\{\mathcal{K}^t\}_{t\in [0,\infty)}$ form a one parameter semigroup indexed by time. All elements of the semigroup share the same eigenfunctions, with varying eigenvalues depending on their parameter value. The generator of the semigroup is \emph{identically} the infinitesimal generator of the SDE. That is, the generator of the sKO semigroup is exactly the evolution operator of the KBE. While this connection has been studied in stochastic analysis since the time of Kolmogorov, this connection is made most explicit in \cite{vcrnjaric2019koopman}. 

With this knowledge, we can construct importance sampling estimators for nonlinear SDEs. Observe that \eqref{eq:qoikoopapprox} 
provides an approximation to the quantity of interest in \eqref{eq:qoi}. Rather than using it directly to estimate the probability of the rare event, we use it to approximate the Doob transform. Observe that
\begin{align}
	\tilde{\Phi}(t,x) = \sum_{i = 1}^N f_i e^{\lambda_i(T-t)} \phi_i(x),
	\label{eq:approxKBEsol}
\end{align}
is an approximate solution to the KBE in \eqref{eq:kbe}. Then we can use the approximate Doob transform, 
\begin{align}
	\tilde{u}(t,x) = \mathbf{B}(x) ^* \frac{\sum_{i = 1}^N f_i e^{\lambda_i(T-t)} \nabla \phi_i(x)}{\sum_{i = 1}^N f_i e^{\lambda_i(T-t)} \phi_i(x)},
	\label{eq:doobeigenfuncs}
\end{align}
to construct a new importance sampling scheme via \eqref{eq:sdealt} and \eqref{eq:girsanov}. Intuitively, if $\tilde{\Phi}$ is a good approximation of the true solution, then the approximate Doob transform will be a good approximation to the true Doob transform, with the guarantee that if there exists a complete set of eigenfunctions, then the estimator will have zero variance as $N\to\infty$. 

In practice, this framework offers considerable flexibility. While \eqref{eq:qoikoopapprox} provides an approximation to the quantity of interest, the errors introduced by truncation, and any additional errors resulting from numerical approximations of the eigenfunctions themselves, cannot easily be characterized. Instead, using the approximation within the Doob transform allows us to resolve these errors through Monte Carlo simulation. Our numerical experiments will demonstrate that even \emph{crude} approximations of a \emph{few} sKO eigenfunctions, where the latter are built from {non-rare trajectories}, can be used to build effective importance sampling methods for rare event probabilities. Moreover, the dynamics of the controlled SDE system \eqref{eq:sdealt} naturally reveal the most likely paths to the rare event.

Next we discuss numerical techniques for approximating the sKO eigenfunctions.

\subsection{Dynamic mode decomposition methods}
\label{sec:dmd}

Dynamic mode decomposition (DMD) methods are a class of data-driven methods that can \revi{approximate} eigenvalues and eigenfunctions of a (stochastic) dynamical system's (stochastic) Koopman operator. The original DMD method was presented in \cite{schmid2010dynamic} as means of model reduction for complex fluid flows. Low-dimensional behavior was extracted from time series data comprising snapshots of high-fidelity fluid dynamics simulations. The connection between DMD and the spectral objects of the Koopman operator was made clear by \cite{rowley2009spectral,tu2013dynamic}, and there has since been considerable interest in developing more effective and efficient DMD methodologies and variants.

DMD methodologies typically use only sample trajectories of the system to \revi{approximate} the Koopman eigenvalues and eigenfunctions, by indirectly approximating the infinitesimal generator \cite{vcrnjaric2019koopman,williams2015data}. 
To avoid introducing errors due to these approximations of the generator, we use the analytical form of the SDE, which in turn provides access to the exact form of the generator of the sKO semigroup. In particular, we \revi{approximate} Koopman eigenfunctions and eigenvalues using a recently developed variant of DMD called \emph{infinitesimal generator extended dynamic mode decomposition} (gEDMD) \cite{klus2019datadriven}. The approach is based on using the stochastic Koopman generator in \eqref{eq:generator} directly. We summarize the main steps in the approach here. 

Fix a set of test points $\{x_i\}_{i = 1}^m$ drawn from a probability measure $\mu$ and a set of twice continuously differentiable basis functions $\{\psi_k(x)\}_{k = 1}^n$.\footnote{In our numerical experiments, we find that collecting test points from sample trajectories tends to produce better results (when validated on separate testing data) than prescribing some arbitrary measure $\mu$.}  Suppose the stochastic process $\{X_t\}$ evolves according to \eqref{eq:sde}. The main idea is to project the action of the Koopman generator onto the basis functions. Following the notation of \cite{klus2019datadriven}, let $\psi(x) = [\psi_1(x), \ldots, \psi_n(x)]^T$, define $\de \psi_k(x) \coloneqq (\mathcal{A}\psi_k)(x)$,
and define
\begin{align}
	d\Psi_X = \begin{bmatrix}
		\de \psi_1(x_1) &\cdots& \de \psi_1(x_m) \\
		\vdots &  \ddots& \vdots \\
		\de \psi_n(x_1) &\cdots& \de \psi_n(x_m) 
	\end{bmatrix} \,\,\,\, 
	\Psi_X = \begin{bmatrix}
		 \psi_1(x_1) &\cdots&  \psi_1(x_m) \\
		\vdots &  \ddots& \vdots \\
		\psi_n(x_1) &\cdots&  \psi_n(x_m) 
	\end{bmatrix}.
\end{align}
Let $K$ be the finite dimensional representation of $\mathcal{A}$. The task is to find the matrix $K\in \mathbb{R}^{n\times n}$ such that the residual $\|\de \Psi_X - K\Psi_X\|_F$ is minimized, where $\| \, \cdot \, \|_F$ is the Frobenius norm. Each column of $K$ is the solution to a least-squares problem, and it can be shown that $K = \de \Psi_X\Psi_X^+$, where $^+$ denotes the pseudoinverse. Furthermore, \cite{klus2019datadriven} shows that as the number of test points $m \to \infty$, this DMD method converges to a Galerkin projection onto the span of the basis functions with respect to $\mu$. Specifically, it is shown that,
\begin{align}
	K = \de \Psi_X \Psi_X^+ = (\de \Psi_X \Psi_X^T)(\Psi_X \Psi_X^T)^+ = \widehat{A}\widehat{G}^+,
		\label{eq:gedmdmat1}
\end{align}
where
\begin{align}
	\widehat{A} = \frac{1}{m}\sum_{i = 1}^m \de \psi(x_i)\psi(x_i)^T, \,\,\,\, \widehat{G} = \frac{1}{m}\sum_{i = 1}^m  \psi(x_i)\psi(x_i)^T. 
	\label{eq:gedmdmat2}
\end{align} 
And as the number of test points goes to infinity, 
\begin{align}
	\lim_{m\to\infty} \widehat{A}_{ij} = \int (\mathcal{A}\psi_i) (x) \psi_j(x)\, \de \mu, \;\;\;\;
	\lim_{m\to\infty} \widehat{G}_{ij} = \int \psi_i(x) \psi_j(x)\, \de \mu.  
\end{align} 
The quality of the approximated eigenfunctions and eigenvalues will depend on the choice of basis functions and test point measure $\mu$. \revi{We discuss our choices of basis functions and $\mu$ within the numerical examples of Section \ref{subsec:numerical} for nonlinear stochastic systems; there, we also describe how we validate the resulting eigenfunction approximations.} \revi{We summarize gEDMD in Algorithm \ref{alg:algorithmedmd}.}

\begin{algorithm}
\small
\SetKwInput{KwInput}{Input}                
 \KwInput{SDE $dX_t = \mathbf{A}(X_t) \de t + \mathbf{B}(X_t) \de W_t$, Basis functions $\{\psi_j(x) \}_{j = 1}^n$, measure $\mu$}
 \KwOut{Stochastic Koopman eigenfunctions $\{\phi_i(x)\}$ and eigenvalues $\{\lambda_i\}$ } 
 \vspace{2pt}
 \begin{algorithmic}[1]
 \STATE Obtain test points $\{x_i\}_{i = 1}^m$ from measure $\mu$
 \STATE Evaluate $\{\psi_j(x)\}_{j = 1}^n$ and $\{\mathcal{A}\psi_j(x)\}_{j = 1}^n$ at the test points
 \STATE Form matrices $\widehat{A}$, $\widehat{G}$, and $K$ in \eqref{eq:gedmdmat1} and \eqref{eq:gedmdmat2}
 \STATE Compute eigenvalues $\{\lambda_i\}_{i = 1}^n$ and eigenvectors of $\{v_i\}_{i = 1}^n$ of $K$
 \STATE Eigenfunctions are $\phi_i(x)= v_i^T \psi(x). $
  \vspace{1pt} 
 \end{algorithmic}
 \caption{\revi{Infinitesimal generator extended dynamic mode decomposition (gEDMD)}}
 \label{alg:algorithmedmd}
\end{algorithm}

\revi{Lastly, we discuss the potential for using other DMD methods. Our primary reason for selecting generator EDMD is that we have direct access to the generator, and would like to exploit it. Other extended dynamic mode decomposition (EDMD) methods could also be applicable \cite{williams2015data}. A drawback of EDMD and its variants, however, is that they require a judiciously chosen basis, which can be difficult to devise in the purely data-driven setting. There are methods such as (stochastic) Hankel DMD \cite{vcrnjaric2019koopman,arbabi2017ergodic} that do \emph{not} require a dictionary of functions, and instead use delay embedding to generate a suitable dictionary. While such methods can accurately approximate the eigenfunctions at the test points, evaluating gradients of the eigenfunctions is more difficult. For our importance sampling approach, we need the ability to evaluate the gradient of the eigenfunctions \emph{cheaply} and at a variety of \emph{non-test} input points. Evaluating the gradients of Hankel DMD eigenfunction approximations is cumbersome, as it either requires interpolation or the solution of adjoint equations. }

\subsection{Approximating observables by eigenfunctions}
\label{sec:regression}

Given a finite collection of sKO eigenfunctions, we can approximate solutions to the KBE, and hence the Doob transform, without having to solve the stochastic optimal control problems associated with existing rare event sampling methods. Computing these sKO eigenfunctions, as described in the previous section, is the first numerical challenge of our approach. The second challenge is to approximate the observable $f$ as a linear combination of the eigenfunctions. We tackle this very simply, using linear regression with a least-squares objective. 

Formulating this regression problem precisely, and ensuring that the results can be used to define an appropriate biasing function via \eqref{eq:doobeigenfuncs}, requires resolving two issues. First is the choice of regression points. We construct the regression problem using the \emph{same} point set used to \revi{approximate} the sKO eigenfunctions, as described in the previous subsection. For rare event simulation, i.e., $f(x) = \mathbbm{1}_E(x)$, properly representing the indicator function demands that some regression points lie inside the event of interest $E$. In our approach, we simulate many trajectories of the original system with different initial conditions throughout the domain and then subsample each trajectory to generate the regression (and EDMD) points. We assume that the user knows where the rare event $E$ lies in state space, but has little idea \emph{how} the system reaches it. To ensure that we have regression points inside $E$, we begin many of the sample trajectories inside the event of interest. In our numerical experiments, for instance, the initial conditions are uniformly spaced over some  subset of the state space that contains a portion of the rare event and the initial condition $x$. (Further details are given in Section~\ref{sec:numericalexamples}.) 

Now suppose that $\{\phi_i\}_{i = 1}^N$ are the computed sKO eigenfunctions, and let $\{x_j\}_{j = 1}^m$ denote the regression points. Let $\mathbf{f} = (f_1, \ldots, f_N) \in \R^N$ be the expansion coefficients in \eqref{eq:approxobservable}, $\mathbf{F}\in\R^m$ be evaluations of $f$ at the regression points, and $\mathbf{C}\in\R^{m\times N}$ be the design matrix with $\mathbf{C}_{ji} = \phi_i(x_j).$ We then solve the least-squares problem,
\begin{align}
    \min_{\mathbf{f}\in\R^N} \|\mathbf{F}-\mathbf{C}\mathbf{f} \|_2^2 \, .
    \label{eq:lsq}
\end{align}

Next, recall that the Doob transform requires the approximate KBE solution \eqref{eq:approxKBEsol} to be strictly positive. This property is not guaranteed by linear regression onto the eigenfunctions. One could add positivity constraints at the regression points to \eqref{eq:lsq}, but instead we correct ``afterwards'' by adding a constant to the approximate KBE solution produced by the regression. This correction does not impact the consistency of the sampling approach, because the constant function is always an sKO eigenfunction. The value of the approximated observable $\tilde{f}(x) = \sum_{i=1}^N f_i \phi_i(x)$ at each of the regression points can be found  by computing $\mathbf{Cf}$. Assuming that the regression points sufficiently sample the relevant parts of the state space, we simply take the minimum of these values, denoted by $-\varepsilon$, and replace the coefficient $f_1$ of the constant sKO eigenfunction with $f_1 + \max(\varepsilon, 0)$. 

It is important to note that adding a positive constant to $\tilde{f}$ will not affect the \emph{direction} of the biasing function $\tilde{u}(t,x)$. The magnitude of $\tilde{u}$ will be diminished, however, since adding a constant increases the magnitude of the denominator in \eqref{eq:doobeigenfuncs}. This correction may thus cause the biasing to be too small to push the state into the rare event. To address this issue, we scale the biasing function by a multiplicative factor $c \ge 1$, to ensure that a sufficient \revi{fraction} of trajectories reach the rare event when performing importance sampling; our final biasing is thus $\tilde{u}(t,x) = c\mathbf{B}(x)^*\nabla \log\tilde{\Phi}(t,x)$. In practice, we adjust $c$ after finding the Doob transform, by simulating small batches of the controlled system with different $c$ values and choosing a value such that a sufficient fraction of samples (e.g., 0.2-0.6) reach the rare event. \revi{If $c$ is chosen  too small, not many samples will reach the rare event and the resulting estimator will have a large variance. If $c$ is chosen too large, then too many samples will be pushed deeply into the rare event, but the resulting weights will be small. Since the estimator is unbiased, this implies that there will be at least one sample with a very large weight, which again implies that the estimator will have a large variance. We have observed that when the Doob transform is well approximated, the best factor $c$ is close to one, meaning that essentially no multiplicative correction is needed. In Section \ref{subsec:1Dexample}, we observe the impact of $c$, and justify our procedure for choosing its value, through an example. }

Our approach is summarized in Algorithm \ref{alg:algorithm}. In the next section, we provide further details on our implementation of the regression method and explore how the numerical choices above affect the performance of importance sampling.

\begin{algorithm}

\small
\SetKwInput{KwInput}{Input}                
 \KwInput{SDE $dX_t = \mathbf{A}(X_t) \de t + \mathbf{B}(X_t) \de W_t$ and observable $f(x)$}
 \KwOut{Approximate Doob transform $\tilde{u}(t,x)$ } 
 \vspace{2pt}
 \begin{algorithmic}[1]
 \STATE Generate test points $\{x_j\}_{j = 1}^m$ from sample trajectories with different initial conditions
 \STATE Apply generator EDMD (Algorithm \ref{alg:algorithmedmd}) to obtain sKO eigenfunctions $\{\phi_i(x)\}_{i = 1}^N$ and eigenvalues $\{\lambda_i\}_{i = 1}^N$. Alternatively, for linear systems, OU eigenfunctions are computed exactly. 
 \STATE Approximate $f(x)\approx \tilde{f}(x) = \sum_{i = 1}^N f_i \phi_i(x)$ via regression 
  \STATE If necessary, increase $f_1$ so that $\tilde{f}(x_j)>0$ for all $j$. 
 \STATE Approximate solution to KBE is $\tilde{\Phi}(t,x) = \sum_{i = 1}^N f_i e^{\lambda_i(T-t)} \phi_i(x) $ 
 \STATE Approximate Doob transform (biasing) is $\tilde{u}(t,x) = c\B(x)^*\nabla \log \tilde{\Phi}(t,x).$ Choose $c$ such that a sufficient number of trajectories reach the rare event. 
  \vspace{1pt} 
 \end{algorithmic}
 \caption{Approximating the Doob transform.}
 \label{alg:algorithm}
\end{algorithm}

\section{Numerical examples}
\label{sec:numericalexamples}

We demonstrate our framework on a series of linear and nonlinear stochastic dynamical systems. The impact of numerical parameters used to construct the biasing is first explored through a simple example involving a one-dimensional Ornstein--Uhlenbeck (OU) process. We then demonstrate the generality of our approach for linear dynamical systems with additive noise, by applying it to a non-normal  linear SDE, a noisy Brownian oscillator, and the stochastic advection-diffusion equation (which is an infinite-dimensional system). We then turn to several nonlinear SDEs, where we show how the approach enables escape from different types of attractors.

The stochastic ODE systems are integrated numerically using a stochastic Runge--Kutta scheme \cite{sarkka2019applied,rossler2010runge}. The stochastic PDE system is integrated using exponential Euler methods \cite{jentzen2009numerical}. Since our importance sampling is unbiased, it suffices to report the variance of the importance sampling weight as seen in \eqref{eq:variance}. Without loss of generality, we will also report the relative error defined in \eqref{eq:varrelerr} with $M = 1$, i.e., the \emph{relative error per sample}.

\subsection{Illustrative one-dimensional SDE}
\label{subsec:1Dexample}
We first consider a simple one-dimensional OU process to illustrate our approach and to highlight numerical challenges that occur in more complex examples as well. Let $X_t\in \R$ evolve according to
\begin{align}
	\begin{dcases}
		\de X_t = -X_t \, \de t  + \sqrt{2}\, \de W_t, \\
	X_0 = 0.
	\end{dcases}
\end{align}
Our goal is to estimate $\rho = \Pb\left(X_T \ge 2 \vert X_0 = 0\right) =  \Ex[\mathbbm{1}_{x>2}(X_T) \vert X_0 = 0]$, where $T = 1$. For this problem, the marginal density at time $T$ can be derived analytically and the exact value of $\rho$ (to five digits) is $1.5745 \times 10^{-2}$. 

\revi{The infinitesimal generator of the system, which is the same as the stochastic Koopman generator}, is 
\begin{align}
	\mathcal{A}\psi = -x \psi ' + \psi'' 
\end{align}
and the associated eigenvalue problem is known as the Hermite differential equation, whose solutions can be found in closed form. The eigenfunctions are the probabilists' Hermite polynomials $\phi_n(x) = \He_n(x)$ with eigenvalues $\lambda_n = -n$ for $n \in \mathbb{N}$. We use least squares regression to find the expansion coefficients in \eqref{eq:approxobservable}. In this case, the eigenfunctions are orthogonal with respect to the standard Gaussian distribution, which implies that an optimal approximation of the indicator function $f(x) = \mathbbm{1}_{x>2}(x)$ in a weighted $L^2(\nu)$ (where $\nu$ is the standard Gaussian measure) sense could be found by integrating the product of the indicator and an eigenfunction over the standard Gaussian measure. More generally, if the diffusion is reversible, then its eigenfunctions will be orthogonal with respect to the invariant distribution of the system \cite{pavliotis2014stochastic}. However, this approach is impractical for higher-dimensional systems, as it would require computing several high dimensional integrals, each of which is sensitive to the rare event. Therefore, to keep this example consistent with the results of the more complicated systems, we perform regression as described in Section~\ref{sec:regression}.

\begin{figure}
  \begin{subfigure}{0.49\textwidth} 
    \includegraphics[width=\linewidth]{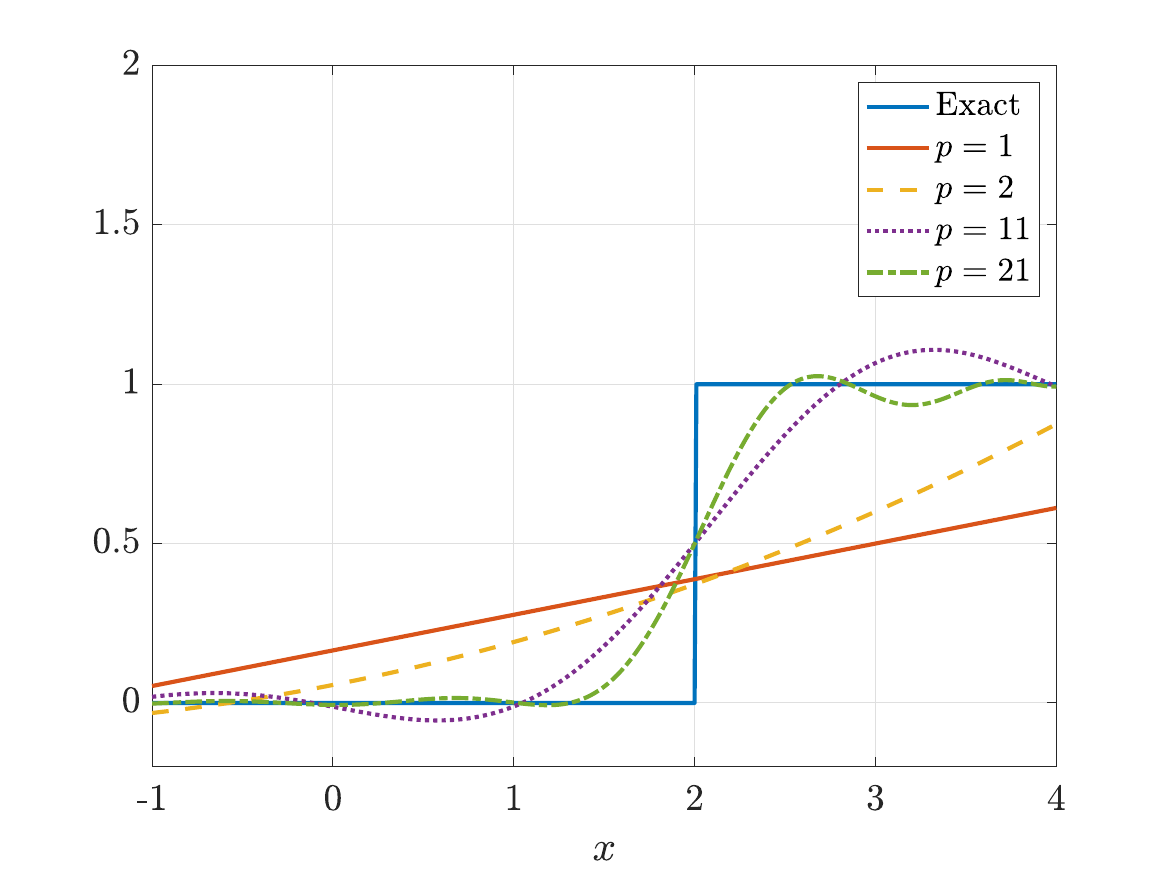}
    \caption{Without positivity constraint.} \label{fig:regression1}
  \end{subfigure}%
  \hspace*{\fill}   
  \begin{subfigure}{0.49\textwidth}
    \includegraphics[width=\linewidth]{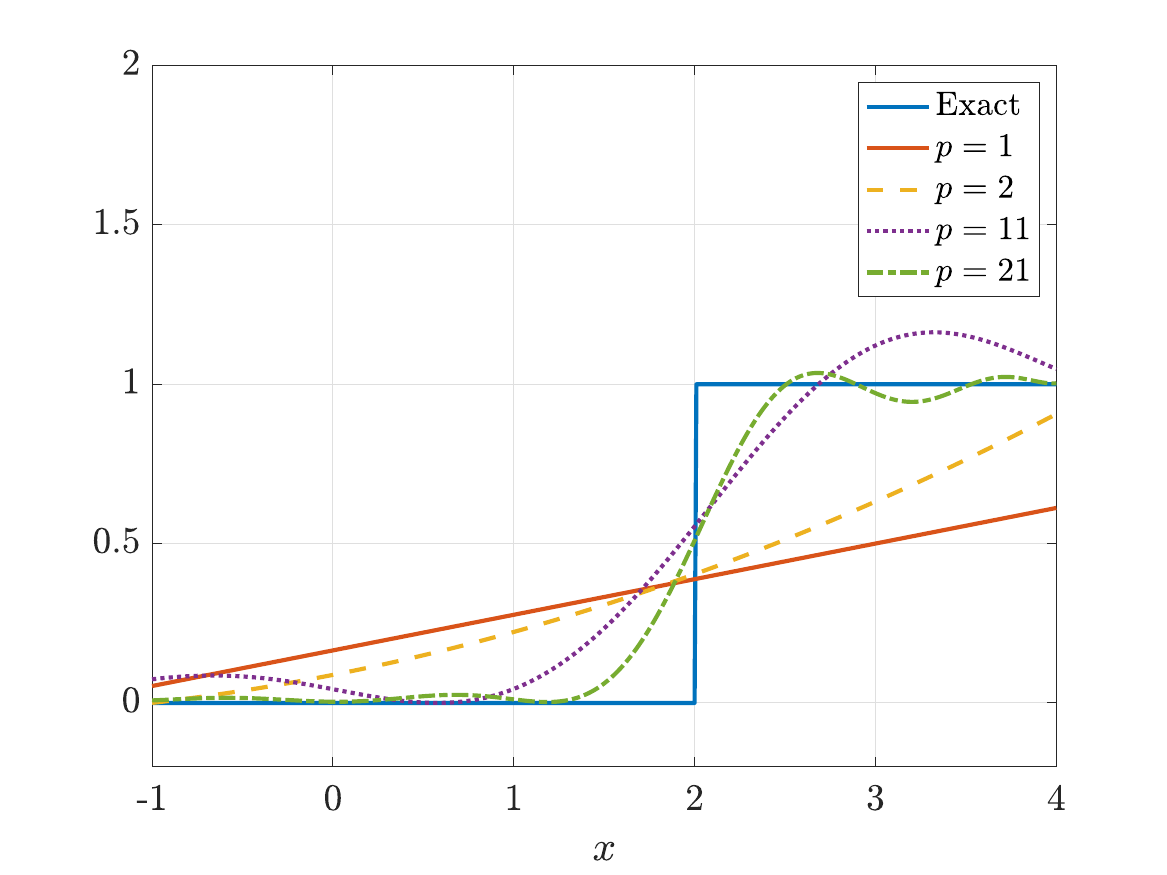}
    \caption{With positivity constraint.} \label{fig:regression2}
  \end{subfigure}%
\caption{Approximating the indicator function.} \label{fig:1}
\end{figure}

We perform our regression with test points drawn from a distribution with more probability mass in the rare event than the invariant distribution. Specifically, we draw $m=50$ independent samples $x_i \sim \mathcal{N}(0,2^2)$, where roughly $15\%$ of points fall inside the region of interest. To mitigate the Gibbs phenomenon, we use a mollified version of the indicator in the regression problem,
    $f(x) =  \frac{1}{2}(1+\tanh(3(x-2)))$. 
The resulting approximations, for polynomial degrees $p = N-1 = 1, \, 2, \, 11,\, 21$, are plotted in Figure \ref{fig:regression1}. Notice that least-squares regression often leads to the approximating function not being strictly positive over the domain. As explained in Section~\ref{sec:regression}, we then add a constant to the approximation such that it is strictly positive. We show the resulting approximations of the indicator function in Figure \ref{fig:regression2}.

Now we use the approximated observable $\tilde{f}$ to build an approximate Doob transform \eqref{eq:doobeigenfuncs} and perform importance sampling via \eqref{eq:sdealt} and \eqref{eq:girsanov}. To account for a diminished biasing magnitude due to positivization, as discussed in Section~\ref{sec:regression}, we multiply the biasing function by a factor $c \geq 1$ to ensure that a sufficient number of trajectories reach the rare event. Below, we will explore how the choice of $c$ impacts the performance of the importance sampling estimator.  

\begin{figure}
    \centering
\begin{subfigure}{0.49\textwidth}
    \includegraphics[width=\linewidth]{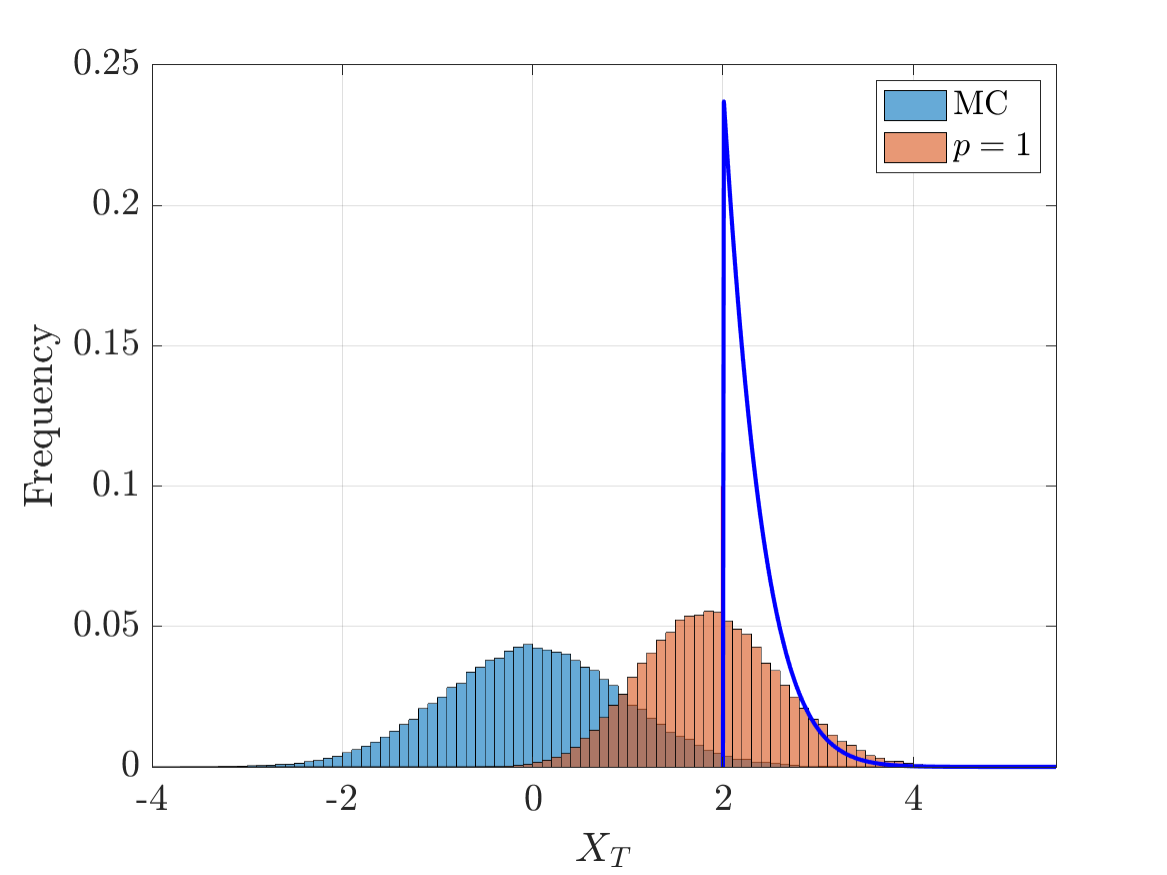}
  \end{subfigure}%
  \hspace*{\fill}   
  \begin{subfigure}{0.49\textwidth}
    \includegraphics[width=\linewidth]{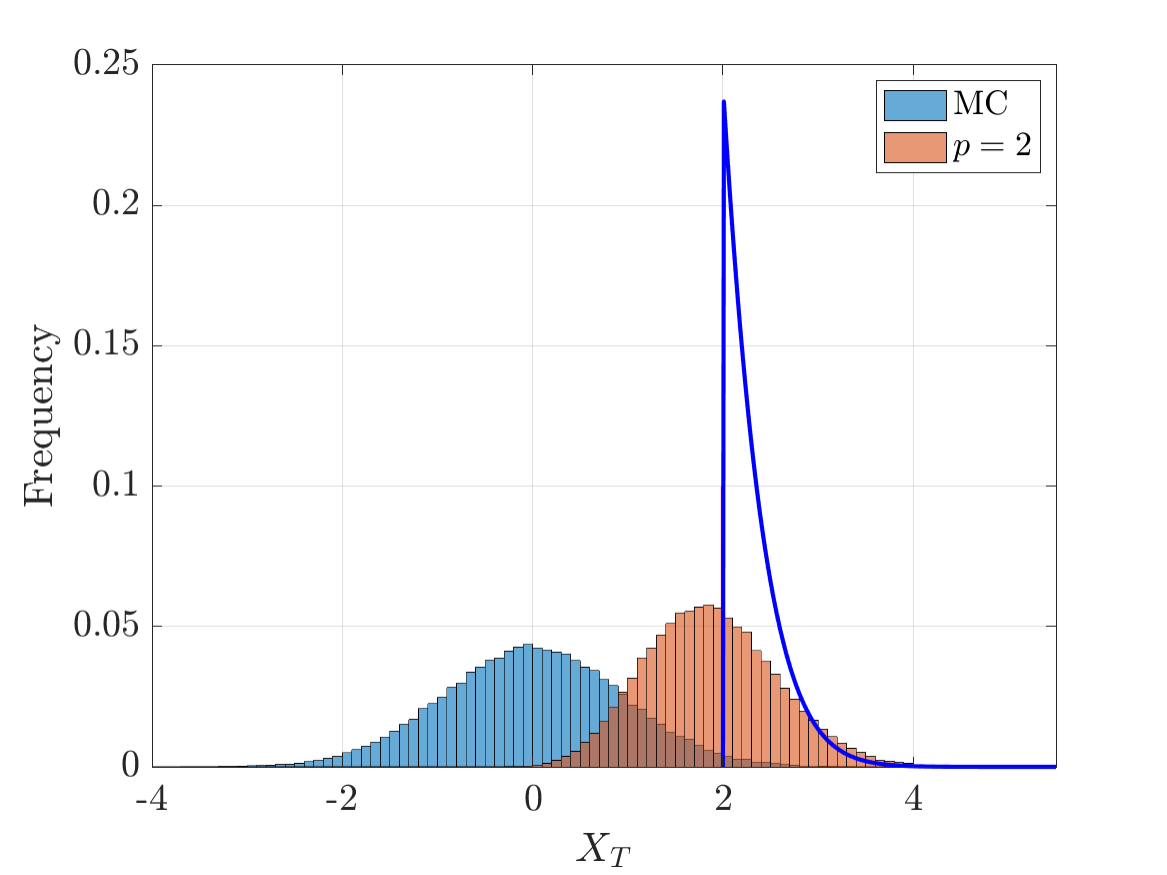}
  \end{subfigure} \\%
  \begin{subfigure}{0.49\textwidth}
    \includegraphics[width=\linewidth]{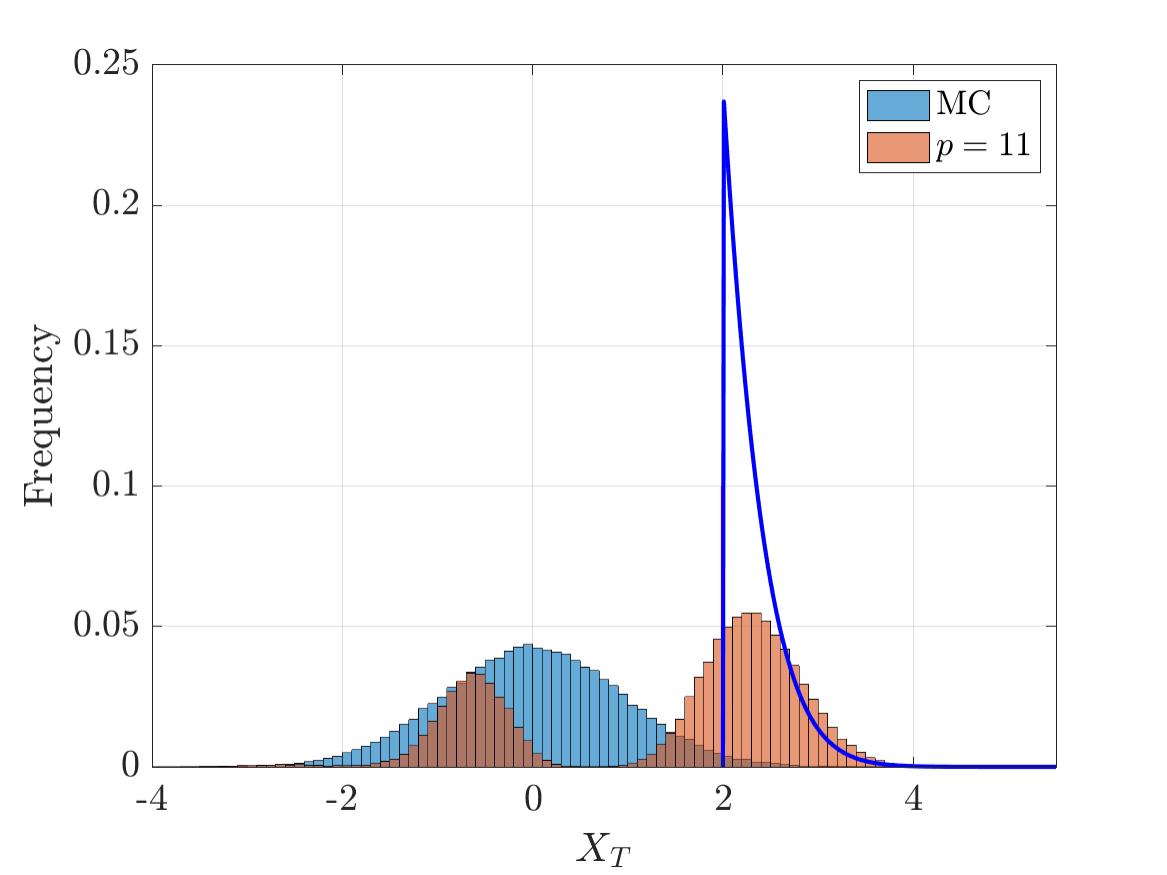}
  \end{subfigure}%
  \hspace*{\fill}   
  \begin{subfigure}{0.49\textwidth}
    \includegraphics[width=\linewidth]{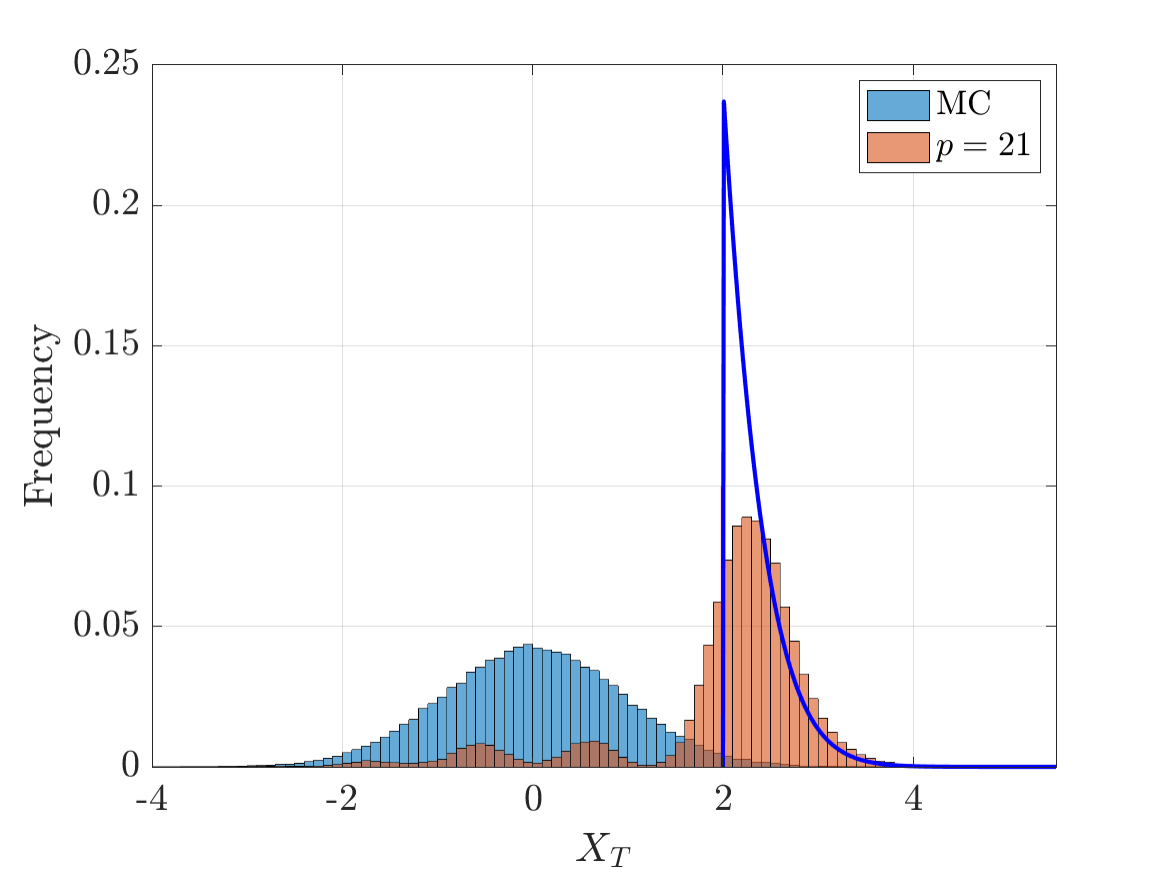}
  \end{subfigure}
    \caption{Sample distribution at time $T=1$ for the one-dimensional OU example. Blue curve is the optimal importance sampling density.}
\label{fig:1Dsampdistr}
\end{figure}

First, Figure \ref{fig:1Dsampdistr} shows the time-$T$ marginal distributions of the biased and unbiased systems, along with the optimal (zero-variance) importance sampling distribution for the expectation of interest. Notice that as the number of eigenfunctions increases, the shape of the histogram tends towards the zero-variance importance sampling density. Table \ref{tab:1Dresult} reports the variance and relative error per sample of the importance sampling estimators resulting from each approximation of $f$. In this simple example, we see that increasing the number of basis functions does not meaningfully increase the efficiency of the estimator. This is likely because increasing the polynomial degree of the approximation leads to more local minima and maxima, causing some sample trajectories to be driven away from the rare event of interest. On the other hand, this result demonstrates how even {a small number of eigenfunctions can significantly improve the efficiency of importance sampling}. For instance, using just two eigenfunctions results in the variance being reduced by a factor of 20 compared to simple Monte Carlo. 

\begin{table}[H]
\centering
\begin{tabular}{llll}
\hline
\multicolumn{1}{|l|}{}   & \multicolumn{1}{l|}{Variance}             & \multicolumn{2}{l|}{Relative error} \\ \hline
\multicolumn{1}{|l|}{Monte Carlo} & \multicolumn{1}{l|}{$1.62\times 10^{-2}$} & \multicolumn{2}{l|}{$8.07$}     \\ \hline
\multicolumn{1}{|l|}{IS $p = 1$} & \multicolumn{1}{l|}{$6.89\times10^{-4}$} & \multicolumn{2}{l|}{$1.67$}      \\ \hline
\multicolumn{1}{|l|}{IS $p = 2$} & \multicolumn{1}{l|}{$7.62\times10^{-4}$} & \multicolumn{2}{l|}{$1.75$}      \\ \hline
\multicolumn{1}{|l|}{IS $p = 11$} & \multicolumn{1}{l|}{$5.56\times10^{-4}$} & \multicolumn{2}{l|}{$1.50$}      \\ \hline
\multicolumn{1}{|l|}{IS $p = 21$} & \multicolumn{1}{l|}{$2.84\times10^{-4}$} & \multicolumn{2}{l|}{$1.07$}      \\ \hline
\multicolumn{4}{l}{$\rho_{\text{true}} = 1.57\times10^{-2}$}                                          
\end{tabular}
\caption{One-dimensional OU example: IS estimator variance with increasing polynomial degree $p$. The multiplier $c$ and offset $\varepsilon$ are tuned.}
\label{tab:1Dresult}
\end{table}

Ideally, the multiplier $c$ should be chosen so that the variance of resulting importance sampling estimator is as small as possible. In Table \ref{tab:multipler} we demonstrate the impact of $c$ on this variance, fixing $p = 1$. The variance of the estimator initially decreases with increasing $c$, up to a threshold beyond which the performance degrades. Intuitively, too small a multiplier results in a larger variance, as too few samples reach the rare event. On the other hand, biasing with a very large multiplier leads to many samples deep in the tails. This implies that the most of resulting weights \eqref{eq:girsanov} will be small. Since the estimator is unbiased, however, a few samples will have very large weights, leading to a large estimator variance overall. In the following numerical examples, we choose $c$ such that roughly 50\% of the resulting samples enter the rare event; this rule of thumb is justified by the trends observed in Table~\ref{tab:multipler}. If, however, most of the weights resulting from a given value of $c$ are very small, and certainly if $c$ is so large such that the relative error is larger than that of simple Monte Carlo, then the value of $c$ should be reduced so that trajectories are not pushed as deeply into the tails.

\begin{table}[H]
\centering
\begin{tabular}{lllllll}
\hline
\multicolumn{1}{|l|}{}   & \multicolumn{1}{l|}{Variance}             & \multicolumn{2}{l|}{Relative error} & \multicolumn{2}{l|}{Proportion in rare event}\\ \hline
\multicolumn{1}{|l|}{Monte Carlo} & \multicolumn{1}{l|}{$1.62\times 10^{-2}$} & \multicolumn{2}{l|}{$8.07$} & \multicolumn{2}{l|}{$0.0157$}     \\ \hline
\multicolumn{1}{|l|}{IS $c = 1$} & \multicolumn{1}{l|}{$6.29\times10^{-3}$} & \multicolumn{2}{l|}{$5.05$}  & \multicolumn{2}{l|}{$0.0432$}     \\ \hline
\multicolumn{1}{|l|}{IS $c = 2$} & \multicolumn{1}{l|}{$2.69\times10^{-3}$} & \multicolumn{2}{l|}{$3.31$}  & \multicolumn{2}{l|}{$0.0912$}     \\ \hline
\multicolumn{1}{|l|}{IS $c = 4$} & \multicolumn{1}{l|}{$8.76\times10^{-4}$} & \multicolumn{2}{l|}{$1.89$}    & \multicolumn{2}{l|}{$0.284$}   \\ \hline
\multicolumn{1}{|l|}{IS $c = 6$} & \multicolumn{1}{l|}{$7.88\times 10^{-4}$} & \multicolumn{2}{l|}{$1.79$} & \multicolumn{2}{l|}{0.558} \\ \hline
\multicolumn{1}{|l|}{IS $c = 8$} & \multicolumn{1}{l|}{$4.27\times10^{-3}$} & \multicolumn{2}{l|}{$4.16$}   & \multicolumn{2}{l|}{$0.789$}     \\ \hline
\multicolumn{1}{|l|}{IS $c= 16$} & \multicolumn{1}{l|}{$3.64\times10^{-1}$} & \multicolumn{2}{l|}{$38.4$}  & \multicolumn{2}{l|}{$0.999$}      \\ \hline
\multicolumn{4}{l}{$\rho_{\text{true}} = 1.57\times10^{-2}$}                                          
\end{tabular}
\caption{One-dimensional OU example: impact of the multiplier on importance sampling performance, with fixed $p = 1$. Rightmost column reports the proportion of sample trajectories terminating in the rare event region.}
\label{tab:multipler}
\end{table}

\subsection{Linear examples} 
We now consider linear SDEs of the form, 
\begin{align}
	\de X_t = \A X_t \, \de t + \B \, \de W_t,
	\label{eq:linearsde}
\end{align}
where $\A \in \R^{d\times d}$ is diagonalizable, has eigenvalues $\{-\sigma_i\}_{i = 1}^d$ with strictly negative real parts, and right and left unit eigenvectors denoted by $\{e_i\}_{i = 1}^d$ and $\{w_i\}_{i = 1}^d$, respectively. Here $\B\in \R^{d\times r}$ and $W_t \in \R^r$ is an $r$-dimensional Brownian motion. We also assume that the left eigenvectors of $\A$ are not in the null space of $\B$. 

For linear stochastic dynamical systems, the sKO eigenfunctions can be found exactly. The generator of the sKO semigroup for linear SDEs is known as the {Ornstein--Uhlenbeck} (OU) operator. In was shown in \cite{metafune2002spectrum} that, under mild conditions, the OU operator has a discrete spectrum in $L^p(\nu)$, where $\nu$ is the stationary measure of the process. Furthermore, \cite{metafune2002spectrum} shows that the eigenfunctions are complete in $L^p(\nu)$ for $p \ge 2$, they have a polynomial structure, and the eigenvalues and eigenfunctions are the same for all $p$. Computing the eigenfunctions, however, presents a separate challenge. It is well known that if the OU operator is self-adjoint, which is the case if $\A$ and $\B$ are symmetric and commute, then the eigenfunctions are the tensorized Hermite polynomials  \cite{pavliotis2014stochastic},
\begin{align}
	\phi_{\bf{n}}(x) = \prod_{k = 1}^d \He_{n_i}\left(\frac{\sqrt{2\sigma_i}}{\|\B^* e_i\|} \langle x,e_i\rangle \right). 
\end{align}
If $\A$ is normal with only complex eigenvalues, and $\B$ commutes with $\A$, then the eigenfunctions are a tensor product of Hermite-Laguerre-It\^o polynomials, first noted in \cite{chen2014eigenfunctions}. Otherwise, one has to consider numerical methods for computing the eigenfunctions  \cite{klus2019datadriven,leen2016eigenfunctions,zhang2020ornstein}. 

\subsubsection{Non-normal dynamics}
We begin with a two-dimensional non-normal system, where 
\begin{align}
	\A = \begin{bmatrix}
		-1 & 0 \\ 1 & -0.3
	\end{bmatrix}, \,\,\, \B = 0.1 \mathbf{I}. 
	\label{eq:nonnormal}
\end{align}
The inner product of the two eigenvectors of $\A$ is $0.8192$, \revi{which reflects the degree of non-normality of the linear system}. The eigenvalues are $\sigma_1 = -0.3$ and $\sigma_2 = -1$, with left eigenvectors $[0.8192, 0.5735]^T$ and $[1,0]^T$, respectively. 
We consider the problem of escaping from a ball of radius $L$ at a fixed finite time $T$,
\begin{align}
	\rho = \Pb\left[\|X_T\| \ge L  \, | \,  X_0 = 0 \right],
\end{align}
where $T = 10$, $L = 0.75$. 
This is a problem of escaping from an attractor, which is well-studied in the computational chemistry literature. Methods such as transition path theory \cite{vanden2006transition} and the string method \cite{weinan2002string} characterize the most likely pathways for trajectories to transition between metastable states. Related methods such as the gentlest ascent dynamics \cite{zhou2010gentlest} find transition paths by pushing the system along the direction of the most slowly decaying \emph{right} eigenvector of the Jacobian at the stable point. In a separate effort, \cite{salins2016rare} justifies using most slowly decaying right eigenvector to construct efficient importance sampling estimators for linear stochastic PDEs, in the presence of a suitable spectral gap. Yet these methods are typically restricted to gradient systems (noisy diffusions on a potential energy surface) or self-adjoint linear systems. Using the Koopman approach, we will demonstrate below that biasing a non-normal linear system along the \emph{left} eigenvector that corresponds to the most slowly decaying mode leads to a significantly better importance sampling estimator.

For the system in \eqref{eq:nonnormal}, one can easily check that the first six eigenfunctions, ordered according to the magnitudes of the eigenvalues and with total polynomial degree up to two, are
\begin{align*}
    &\phi_{0} (x) = 1 ,\,\,\,\, &\lambda_0 = 0 \\
    &\phi_{1}(x) = \sqrt{200\sigma_1}\langle x,w_1\rangle \,\,\,\, & \lambda_1 = -0.3 \\
    &\phi_{2}(x) = \sqrt{200\sigma_2}\langle x,w_2 \rangle \,\,\,\, & \lambda_2 = -1 \\
    &\phi_3(x) = {200\sigma_1} \langle x,w_1 \rangle ^2 -1 \,\,\,\, & \lambda_3 = -0.6 \\
    &\phi_4(x) = 200\sqrt{\sigma_1\sigma_2}\langle x,w_1\rangle \langle x,w_2\rangle - 2\frac{\sqrt{\sigma_1\sigma_2}}{\sigma_1+\sigma_2} \langle w_1,w_2\rangle \,\,\,\, & \lambda_4 = -1.3 \\
    &\phi_5(x) = 200\sigma_2\langle x,w_2 \rangle ^2 - 1 \,\,\,\, & \lambda_5 = -2 
\end{align*}
where $w_1$ and $w_2$ are left eigenvectors of $\A$. The function of interest is an indicator on the ball of radius $0.75$ centered at the origin, which is projected onto the set of eigenfunctions. Since the indicator is an even function along the direction of any solitary left eigenvector, we can omit eigenfunctions with odd polynomial degree prior to projection. Thus, the indicator function over the ball is projected onto the span of $\{\phi_0,\phi_3, \phi_4, \phi_5\}$. 

\begin{figure}
\centering
	\includegraphics[width = 0.47\textwidth]{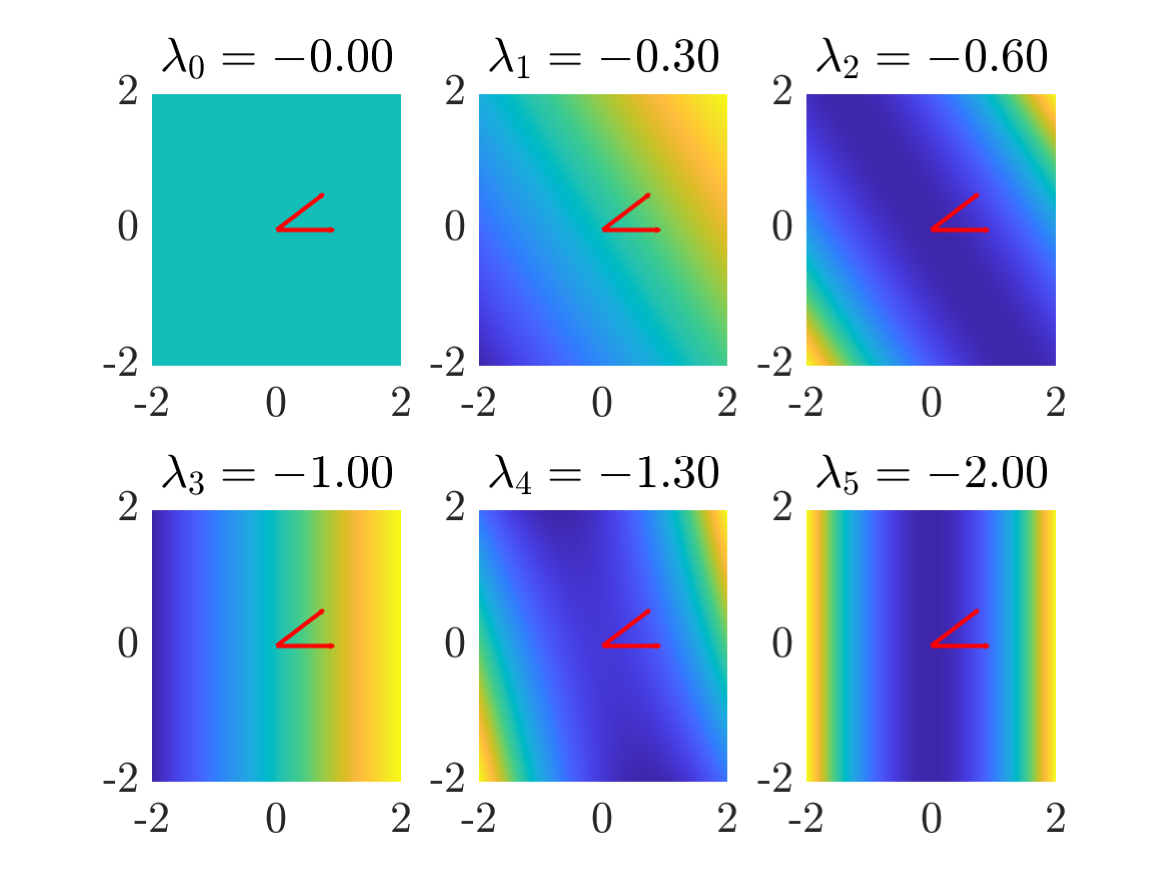}
	\includegraphics[width = 0.52\textwidth]{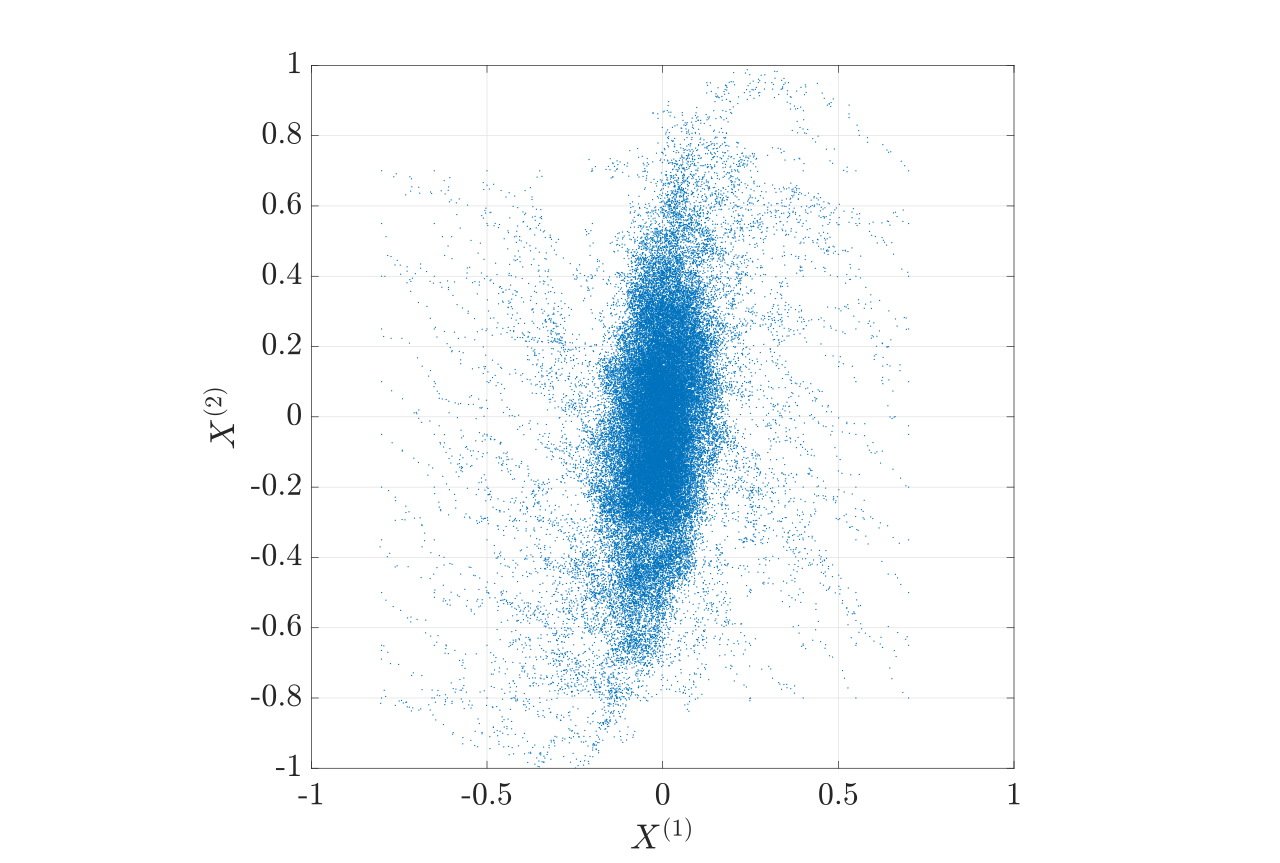}
	\caption{Left: Eigenfunctions of the non-normal SDE. Red vectors illustrate the left eigenvectors of $\A$. Right: regression points generated from sample trajectories of the unbiased system. }
	\label{fig:nonnorm_eigfunc}
\end{figure}

\begin{figure}
\centering
	\includegraphics[width = 0.32\textwidth]{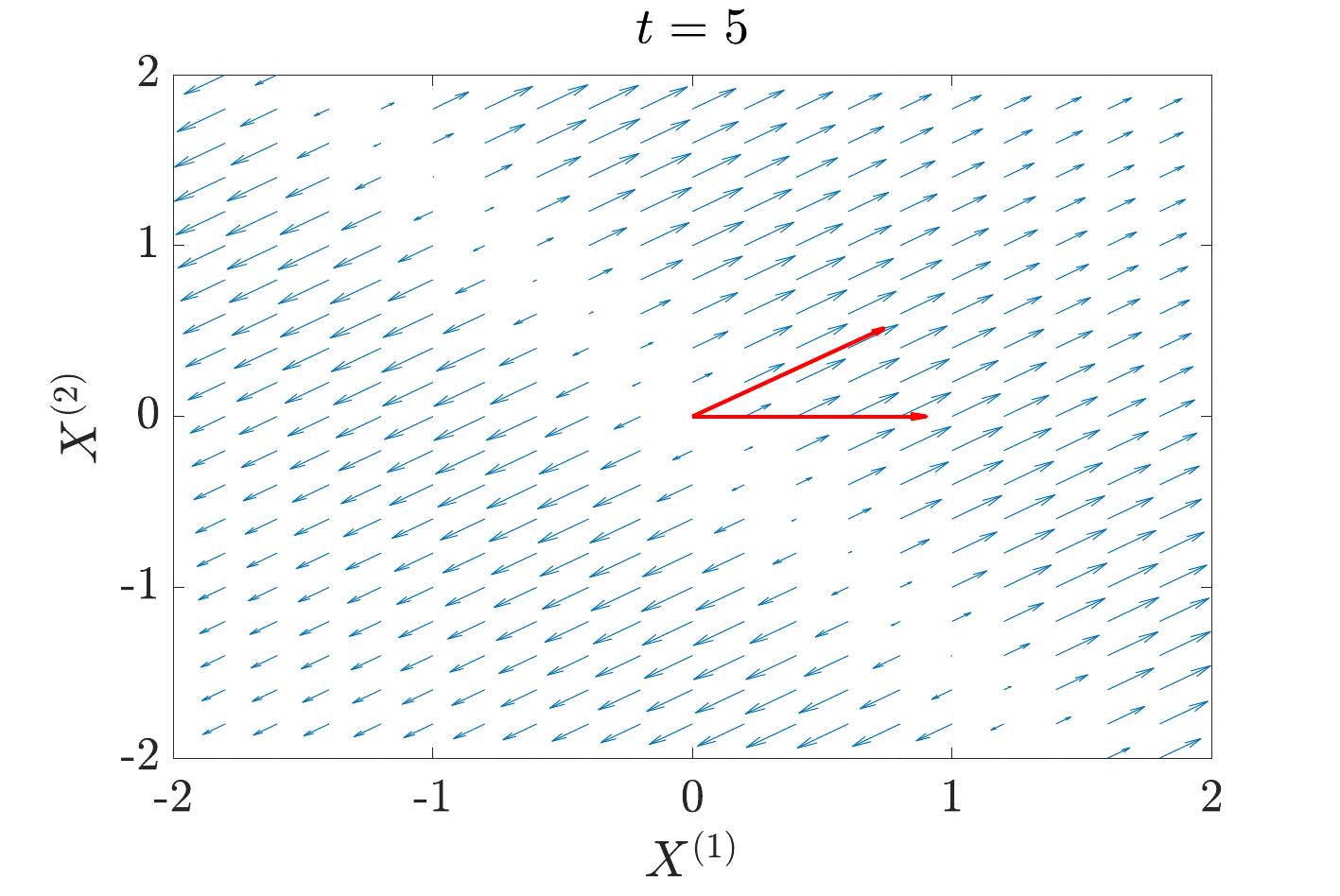}
	\includegraphics[width = 0.32\textwidth]{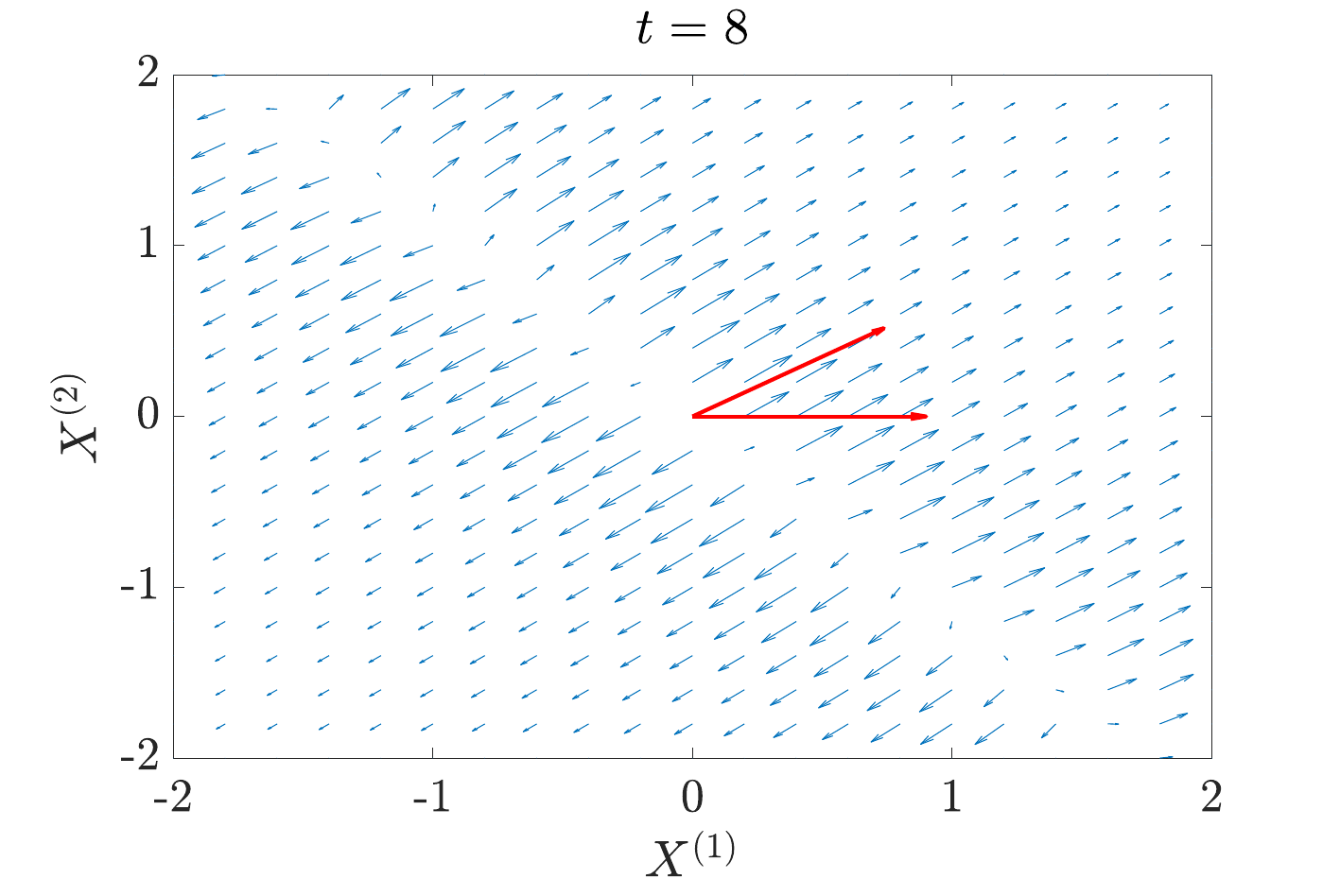}
	\includegraphics[width = 0.32\textwidth]{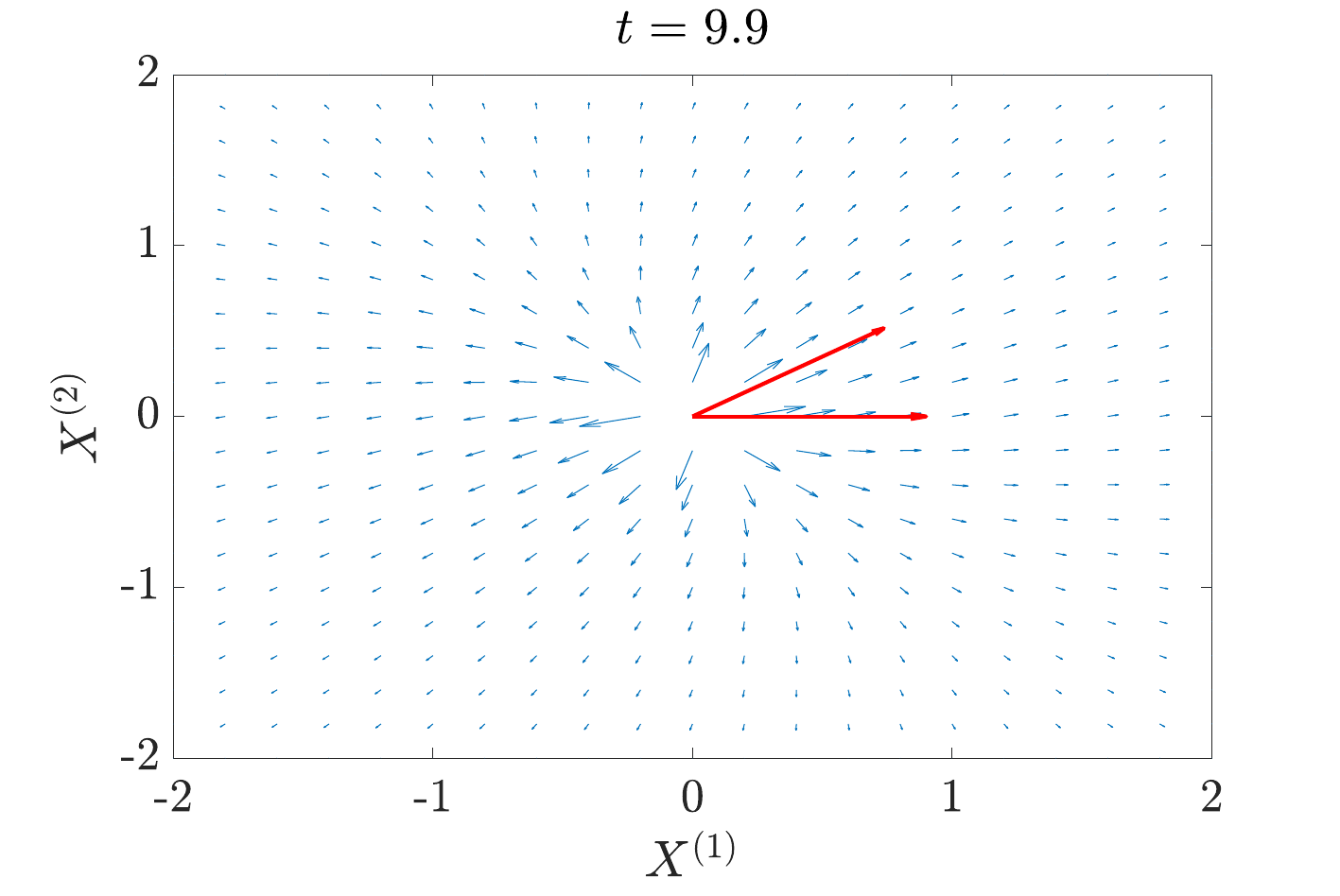}
	\caption{Vector fields of the biasing for the non-normal linear system at different times. Red vectors illustrate the left eigenvectors of $\A$. Note that the lengths of the vectors for a given time are plotted relative to each other and are not comparable for different times. } 
	\label{fig:vectorfield}
\end{figure}

\begin{figure}
\centering
	\includegraphics[width = 0.37\textwidth]{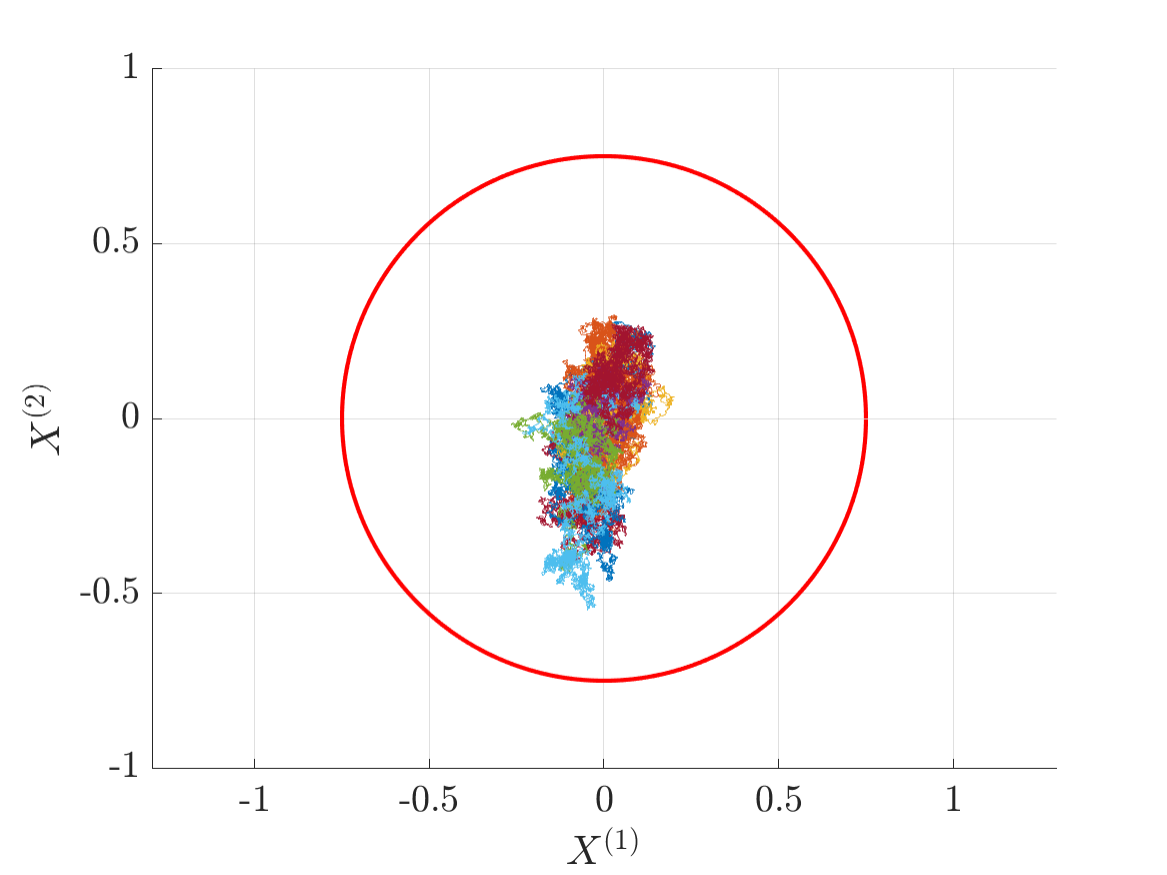}
	\includegraphics[width = 0.37\textwidth]{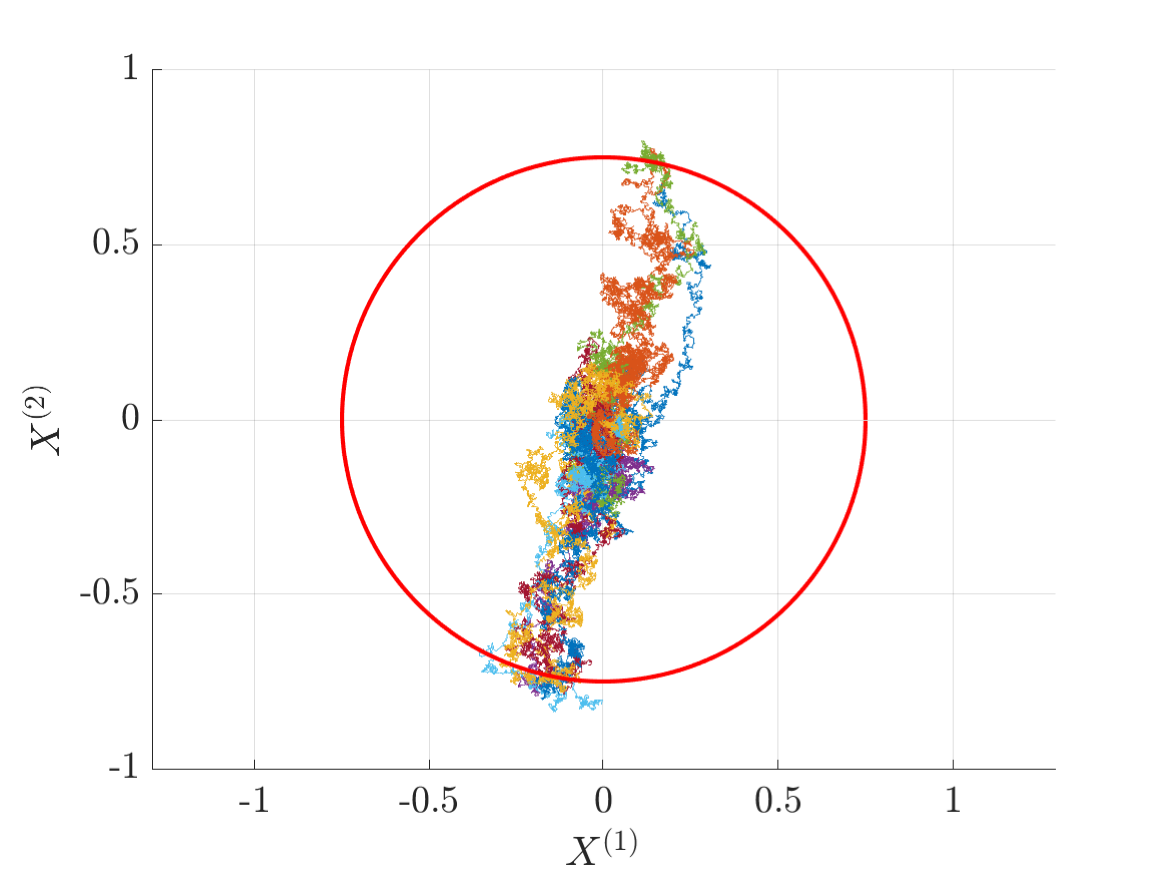}
	\caption{Samples of the nominal and biased trajectories of the non-normal linear system. Red circle denotes the boundary of the rare event.}
	\label{fig:nonnorm_samps}
\end{figure}

\begin{figure}
	\centering	
	\includegraphics[width = 0.5 \textwidth]{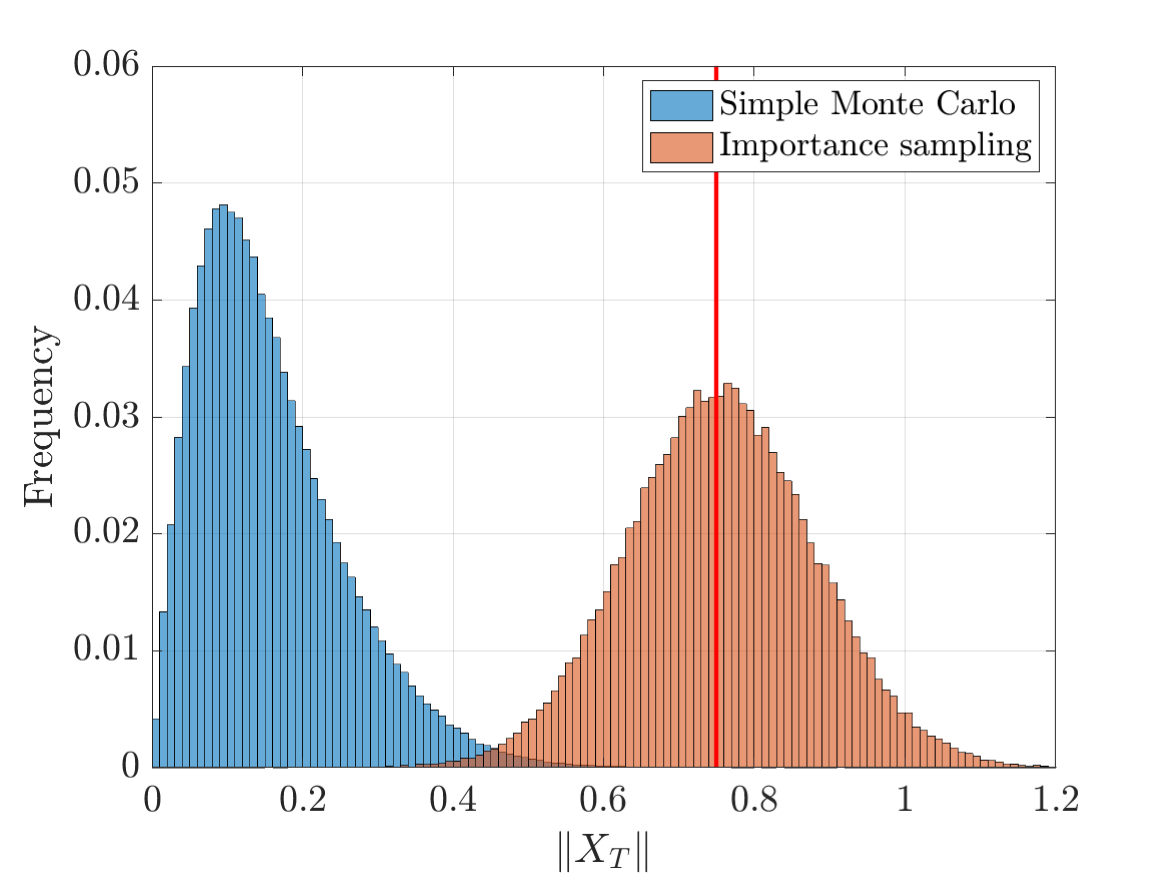}
	\caption{Distribution of the norm of $X_T$ for simple Monte Carlo and importance sampling of the linear non-normal system. Red line denotes the boundary of the rare event region.}
	\label{fig:nonnorm_distr}
\end{figure}

We plot the eigenfunctions in Figure \ref{fig:nonnorm_eigfunc} and highlight the left eigenvector directions in red. To generate the regression points that are used to approximate the indicator function as a linear combination of eigenfunctions, we simulate 121 independent trajectories of length $T$, beginning with uniformly spaced initial conditions on $[-0.8,0.8]^2$. The state is extracted at time intervals of $\Delta t = 0.02$, and the resulting 60621 points are shown in Figure \ref{fig:nonnorm_eigfunc}. \revi{In this example we found 
\begin{align}
    \Phi(t,x)= 0.035 + 0.0342 \phi_3(x) e^{\lambda_3 (T-t)} -0.0323 \phi_4(x) e^{\lambda_4(T-t)} + 0.0092 \phi_5(x) e^{\lambda_5(T-t)}
    \label{eq:nonnormalkbesol}
\end{align}
with multiplicative factor $c = 7$.} Figure \ref{fig:vectorfield} then shows the vector fields produced by the resulting biasing function at $t \in \{5, 8, 9.9\}$. Notice that the biasing pushes in the direction of the slowest-decaying left eigenvector for most of the simulation period $[0,T]$, until the end of the interval, when $T-t$ becomes small and the biasing \eqref{eq:doobeigenfuncs} begins to push in all directions away from the attracting point. In Figure \ref{fig:nonnorm_samps}, we show sample trajectories of the unbiased and biased systems. Notice that the exit directions do not generally align with any eigenvector directions. Performance of the importance sampling estimator is summarized in Table \ref{tab:nonnormal}, where we observe that the variance is reduced by a factor of over 6000. Figure \ref{fig:nonnorm_distr} shows the histogram of the norm of the state at time $T$ for simple Monte Carlo and importance sampling. Notice also that far more samples reach the rare event when importance sampling is applied.

\revi{To show that method works well with larger values of $T$, we also consider the case where $T = 50$ and apply the same biasing scheme, i.e., using \eqref{eq:nonnormalkbesol} but with the new $T$ value. For this case, the estimator performs similarly well and we see that the variance is reduced by a factor of over 3000. The results are summarized in Table \ref{tab:nonnormal2}. The quality is maintained mainly due to the nature of the problem we are solving. We only consider the probability that the state is in the rare event at a particular time $T$, rather than being in the event at any time before $T$. This means that the biasing function need not be very large until close to time $T$. This is reflected in the form of the Doob transform---the biasing function is initially small, but grows exponentially as $t$ approaches $T$. }

\begin{figure}
	\centering
	\includegraphics[width = 0.5\textwidth]{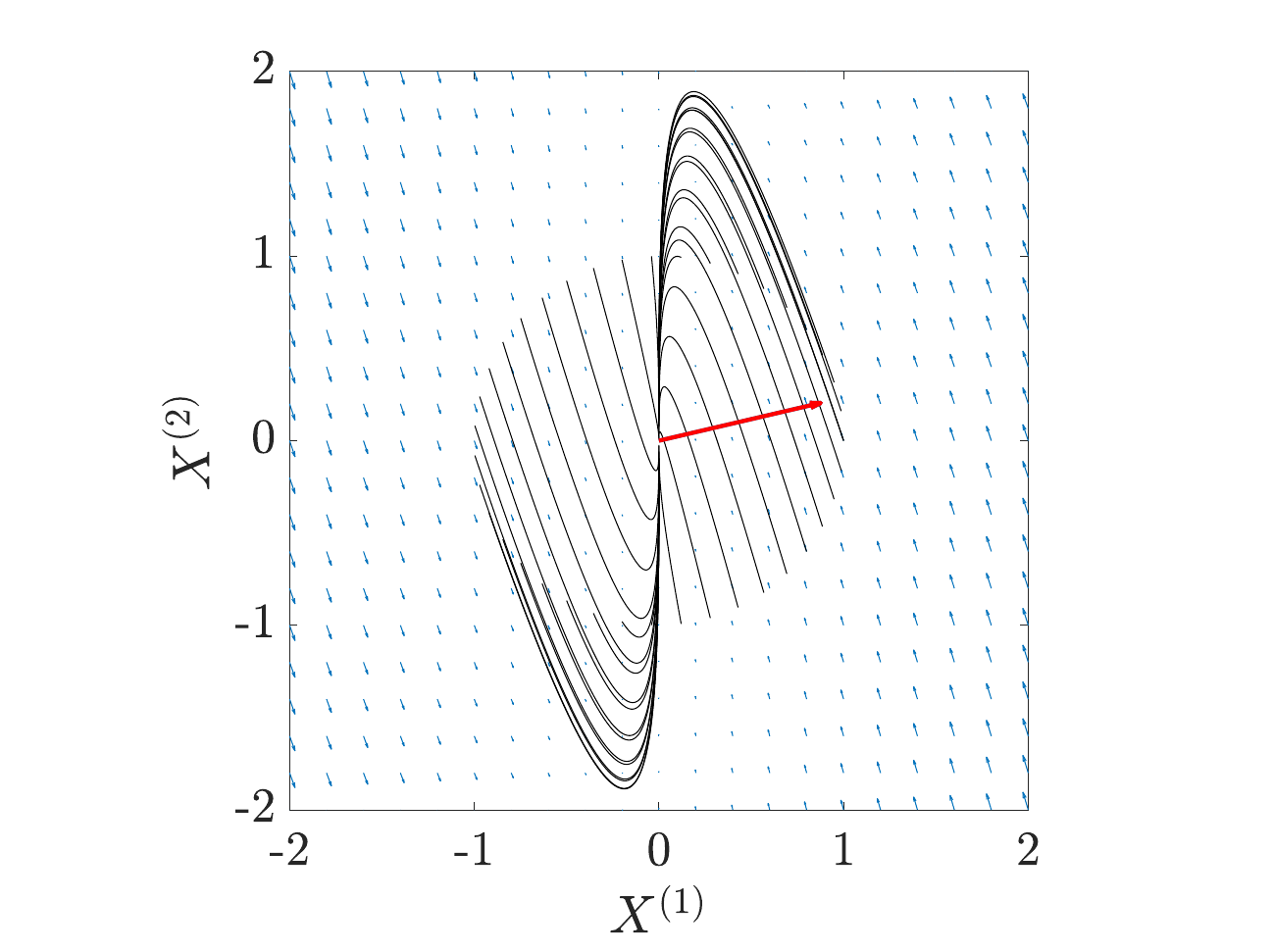}
	\caption{Example phase portrait of a highly non-normal system. The red vector points in the direction of the left eigenvector with the least negative eigenvalue. Notice that initial conditions that lie on the line defined by this eigenvector will initially experience transient growth before decaying to the origin. }
	\label{fig:portraitnonnormal}
\end{figure}

The effectiveness of biasing in the direction of the slowest-decaying left eigenvector can be explained intuitively by considering the phase portrait of a deterministic non-normal linear dynamical system. In Figure \ref{fig:portraitnonnormal}, we plot trajectories of a highly non-normal stable linear system 
with initial conditions on the unit circle. We also plot the left eigenvector with the least negative eigenvalue. First, note that there are initial conditions for which the norm of the state initially grows before decaying towards zero; this is a hallmark of highly non-normal systems. Second, notice that pushing outwards in the direction of the left eigenvector naturally exploits the system's transient growth to move the state even further from the attracting point at the origin. In non-normal systems, left and right eigenvectors corresponding to different eigenvalues are orthogonal. Therefore, the slowest-decaying left eigenvector is orthogonal to the (fast-decaying) manifold spanned by all but the slowest-decaying right eigenvector. When pushing in the direction of the left eigenvector, trajectories are driven away from the attracting point by the fast-decaying manifold of the dynamical system. The left eigenvector direction thus harnesses the system's underlying dynamics to reach the rare event region with the least biasing effort. 

\begin{table}[H]
\centering
\begin{tabular}{llll}
\hline
\multicolumn{1}{|l|}{}   & \multicolumn{1}{l|}{Variance}             & \multicolumn{2}{l|}{Relative error} \\ \hline
\multicolumn{1}{|l|}{Monte Carlo} & \multicolumn{1}{l|}{$1.64\times 10^{-5}$} & \multicolumn{2}{l|}{$246.8$}     \\ \hline
\multicolumn{1}{|l|}{Importance sampling} & \multicolumn{1}{l|}{$2.72\times10^{-9}$} & \multicolumn{2}{l|}{$3.18$}      \\ \hline
\multicolumn{4}{l}{$\rho_{\text{true}} = 1.64\times10^{-5}$}                                         
\end{tabular}
\caption{Importance sampling performance for the SDE with non-normal dynamics. Here, $T = 10$.}
\label{tab:nonnormal}
\end{table}
\begin{table}[H]
\centering
\begin{tabular}{llll}
\hline
\multicolumn{1}{|l|}{}   & \multicolumn{1}{l|}{Variance}             & \multicolumn{2}{l|}{Relative error} \\ \hline
\multicolumn{1}{|l|}{Monte Carlo} & \multicolumn{1}{l|}{$1.74\times 10^{-5}$} & \multicolumn{2}{l|}{$239.6$}     \\ \hline
\multicolumn{1}{|l|}{Importance sampling} & \multicolumn{1}{l|}{$5.60\times10^{-9}$} & \multicolumn{2}{l|}{$4.30$}      \\ \hline
\multicolumn{4}{l}{$\rho_{\text{true}} =1.74 \times10^{-5}$}                                         
\end{tabular}
\caption{Importance sampling performance for the SDE with non-normal dynamics. Here, $T = 50$.}
\label{tab:nonnormal2}
\end{table}

\subsubsection{Stochastically-forced damped harmonic oscillator}
Next we consider a damped harmonic oscillator forced by Brownian motion: 
\begin{align}
	\begin{dcases}
		\ddot{x} + 2 \zeta \omega_0 \dot{x} + \omega_0^2 x = \dot{W}_t \\
		x(0) = x_0 , \, \dot{x}(0) = 0 \, .
	\end{dcases} 
\end{align}
This example will show that our framework works well with complex eigenvalues and \emph{rank-deficient noise}. The oscillator can be put in the form of \eqref{eq:linearsde} with 
\begin{align}
	\A = \begin{bmatrix}
		0 & 1 \\ - \omega_0^2 & -2\zeta \omega_0
	\end{bmatrix} ,\,\,\, \B = \begin{bmatrix}
		0 \\ 1
	\end{bmatrix}. 
\end{align}
We compute $\Pb\left[ \vert x(T) \vert  > L \, \vert \, x(0) = \dot{x}(0) = 0 \right]$, i.e., the probability that the position of the oscillator exceeds some threshold by a fixed time given that it was initially at rest. We set $\omega_0 = 1$, $\zeta = 0.5$, $L = 3$, and $T = 10$.

\begin{figure}
\centering
	\includegraphics[width = 0.49\textwidth]{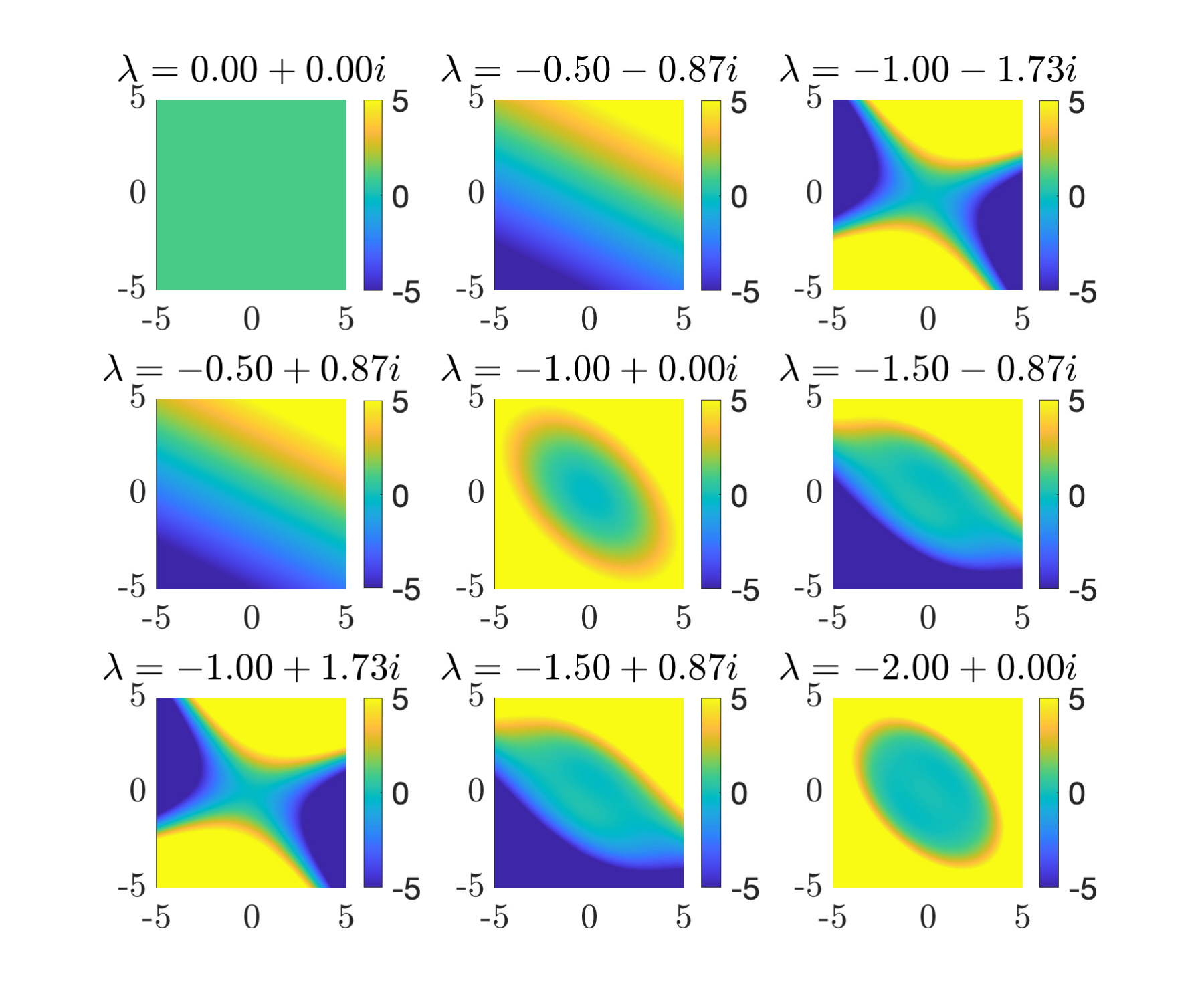}
	\includegraphics[width = 0.49\textwidth]{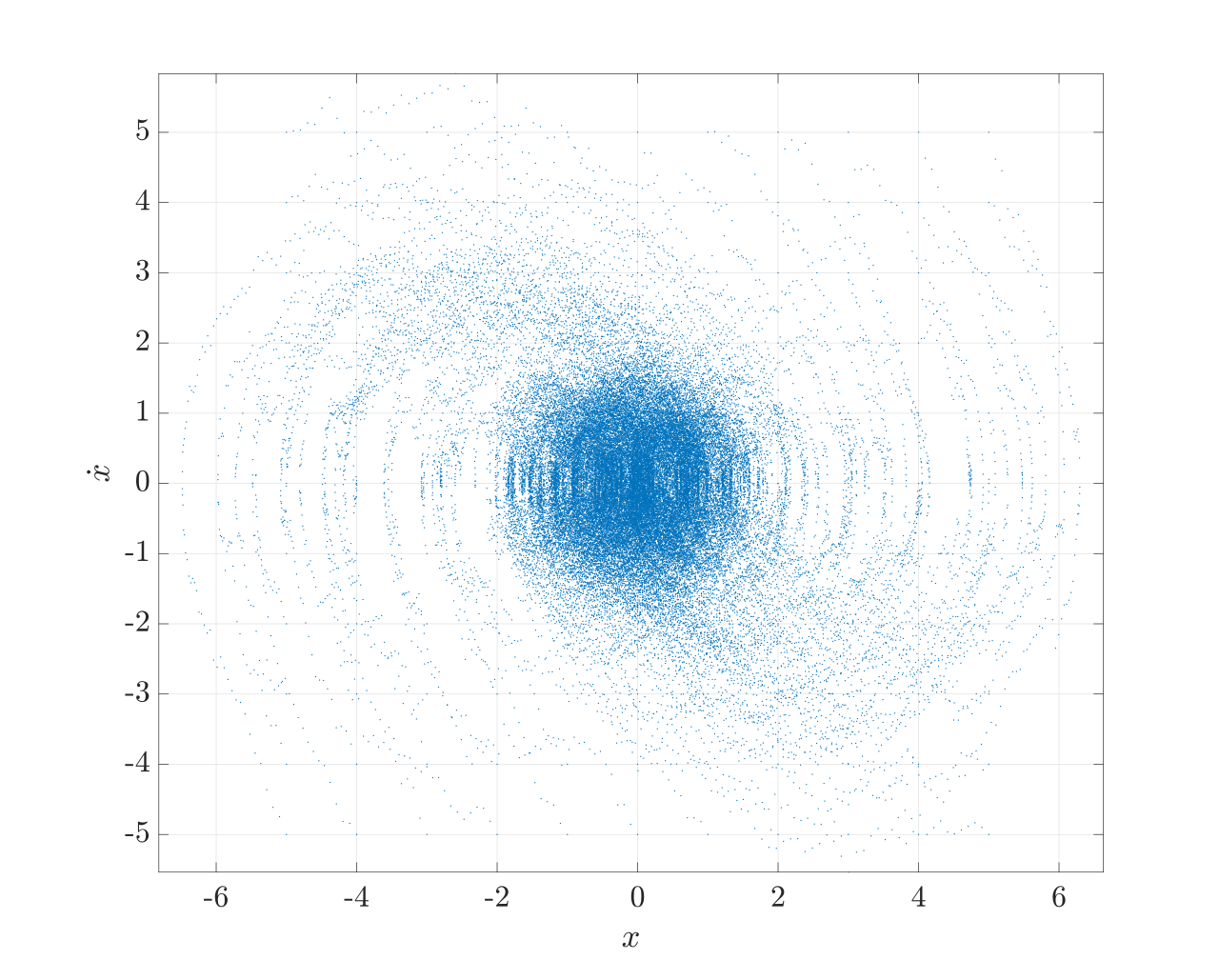}
	\caption{Left: Exact eigenfunctions of the Brownian oscillator. \revi{Only the real part of each eigenfunction is plotted.} Right: regression points based on sample trajectories.  }
	\label{fig:brownosc_eigfuncs}
\end{figure}

\begin{figure}
	\centering	
	\includegraphics[width = 0.75\textwidth]{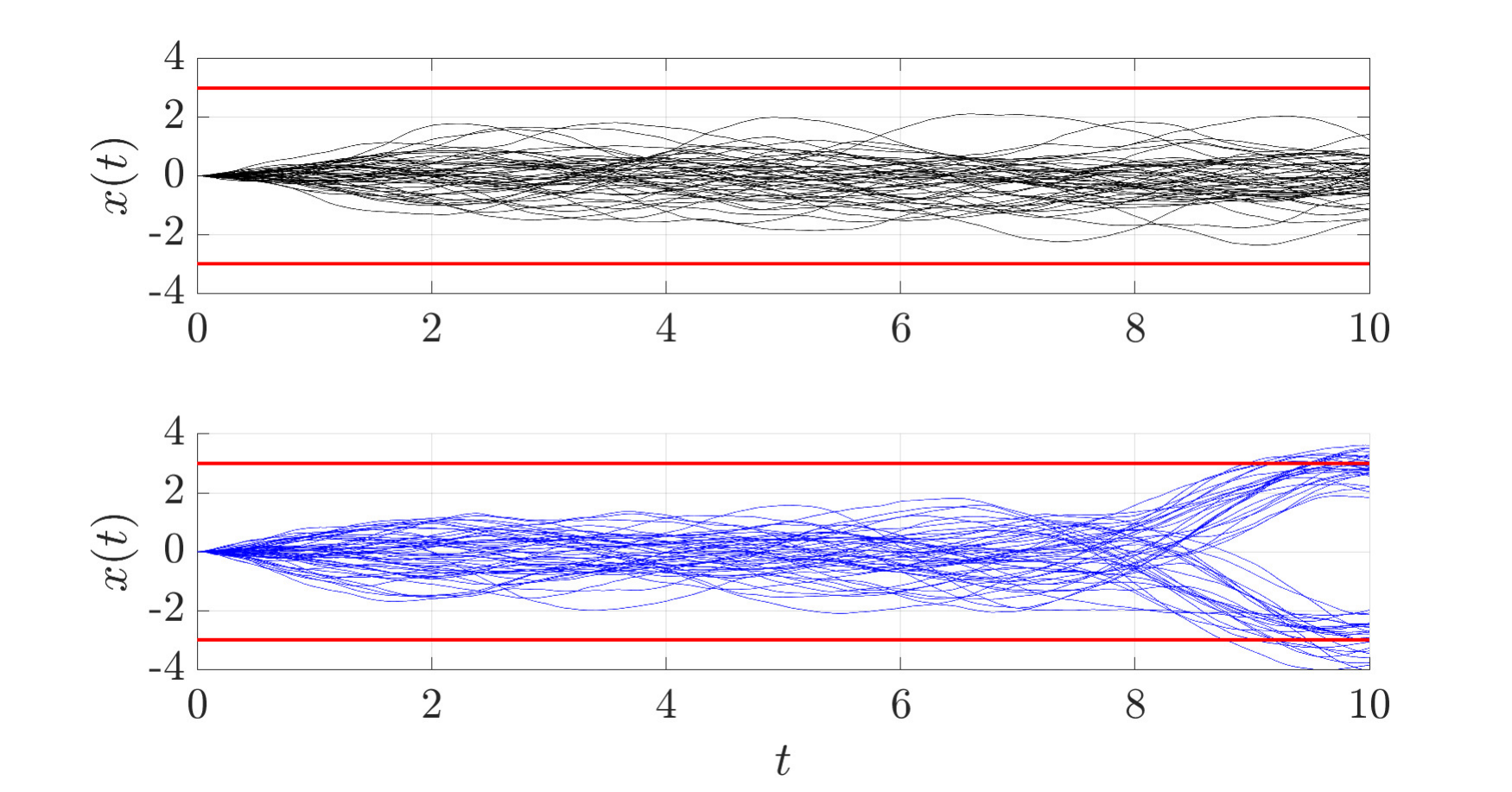}
	\caption{Sample paths of the unbiased and biased Brownian oscillator.}
	\label{fig:brownosc_samples}
\end{figure}

To the best of the authors' knowledge, for the oscillatory case, all rare event simulation algorithms require solving an associated optimal control problem. Here, we instead project an indicator function dependent on the first component of the state, $\mathbbm{1}_{|x|>3}$, onto the first nine sKO eigenfunctions. \revi{The eigenfunctions are expressed as linear combinations of the Hermite--Laguerre--It\^o polynomials; see \cite{chen2014eigenfunctions,zhang2020ornstein}. }
We plot the real parts of these eigenfunctions in Figure \ref{fig:brownosc_eigfuncs}. Regression points are generated by simulating 121 independent trajectories of length $T$ with uniformly spaced initial conditions on $[-5,5]^2$ and extracting the state every $\Delta t = 0.02$ time units. In this example, the constant $c = 6$. In Figure \ref{fig:brownosc_samples}, we show sample trajectories of the unbiased and biased systems. Notice that the impact of  the biasing only seems prominent towards the end of the simulation, e.g., from $t = 8$ onwards. Intuitively, this is because the system has an attracting point at zero, and since we want samples to escape at the end of the simulation, it is not advantageous to bias early. 

\begin{figure}
	\centering
	\includegraphics[width = 0.5\textwidth]{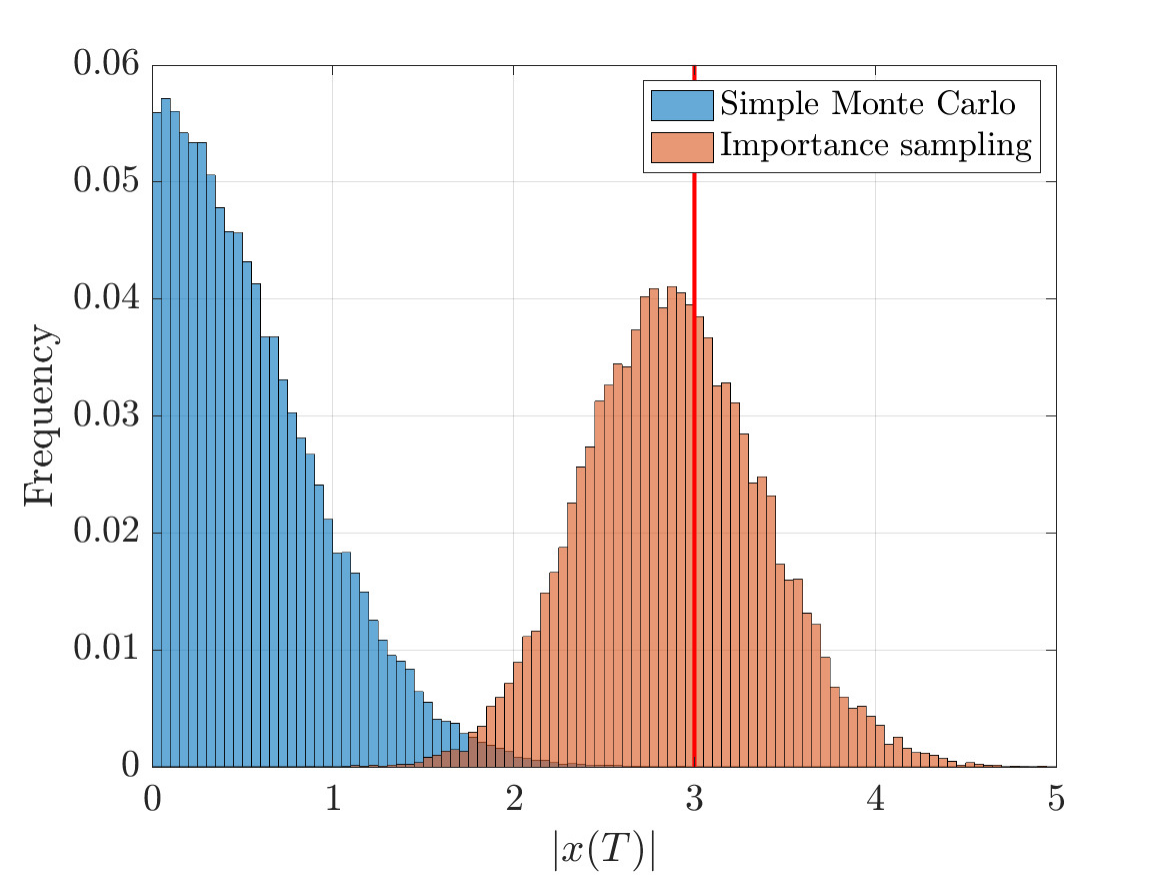}
	\caption{Histograms of $|x(T)|$ obtained using simple Monte Carlo and importance sampling for the Brownian oscillator.}
	\label{fig:brown_distr}
\end{figure}

In Figure \ref{fig:brown_distr} we show histograms of the absolute values of the position of the two systems at time $T$. The estimator performance is summarized in Table \ref{tab:brownosc}, where we observe that biasing reduces the variance by a factor of nearly 5000. 

\begin{table}[H]
\centering
\begin{tabular}{llll}
\hline
\multicolumn{1}{|l|}{}   & \multicolumn{1}{l|}{Variance}             & \multicolumn{2}{l|}{Relative error} \\ \hline
\multicolumn{1}{|l|}{Monte Carlo} & \multicolumn{1}{l|}{$2.28\times 10^{-5}$} & \multicolumn{2}{l|}{$209.5$}     \\ \hline
\multicolumn{1}{|l|}{Importance sampling} & \multicolumn{1}{l|}{$5.10\times10^{-9}$} & \multicolumn{2}{l|}{$3.13$}      \\ \hline
\multicolumn{4}{l}{$\rho_{\text{true}} = 2.28\times10^{-5}$}                                         
\end{tabular}
\caption{Importance sampling performance for the Brownian oscillator.}
\label{tab:brownosc}
\end{table}

\subsubsection{Stochastic advection-diffusion} 
\label{sec:advdiff}
The stochastic advection-diffusion equation is an \emph{infinite}-dimensional non-normal linear system. 
We have, 
\begin{align}
	\begin{dcases}
		v_t = b v_x + \alpha v_{xx}  + \sqrt{\epsilon} \eta \\
		v(t,0) = v(t,1) = 0 \\
		v(0,x) = 0,
	\end{dcases}
\end{align}
where $\eta$ is space-time white noise. Following the approach in \cite{salins2016rare}, this system can be converted into the form of \eqref{eq:linearsde}, where $\mathbf{A}v = bv_x + \alpha v_{xx}$ acts on the space of $L^2$ functions over $x \in [0,1]$ that satisfy the above boundary conditions, $\mathbf{B}$ is the identity map, and $W_t$ is a cylindrical Wiener process. The system is discretized using an exponential Euler method \cite{jentzen2009numerical}. Details about the numerical method used to simulate this process are described in Appendix \ref{sec:euler}. We estimate $\Pb\left[\|v(T, \cdot)\|_{L^2([0,1])} \ge L \right]$ given that the system initially started at $v(t=0, x ) = 0$. We have $b = 1$, $\alpha = 0.1$, $\epsilon = 1$, $T = 10$, and $L = 2.5$. 

We compute the biasing for the system based on the SDE that arises from the discretized version of the stochastic advection-diffusion equation. In this case, we use only two eigenfunctionals: the constant functional and the second order eigenfunctional $\phi_2(v) = \sqrt{2\mu_1}\langle v,w_1\rangle ^2 -1$, where $w_1$ is the eigenfunction of the $L^2$-adjoint of $\A$, $-\sigma_1$ is the leading eigenvalue of $\A$, and $\langle u,v\rangle = \int_0^1 uv \, dx$. \revi{We measure the degree of non-normality of the system by either looking at the P{\'e}clet number, which is equal to $b/\alpha = 10$, or the inner product between the first two eigenfunctions of the advection-diffusion operator, which is equal to $0.9147$. In this example, we obtain 
\begin{align}
    \Phi(t,v) = 0.1434 \phi_2(v) + 1.1434
\end{align}
with $c = 20$.
}

\begin{figure}
	\centering
	\includegraphics[width = 0.5\textwidth]{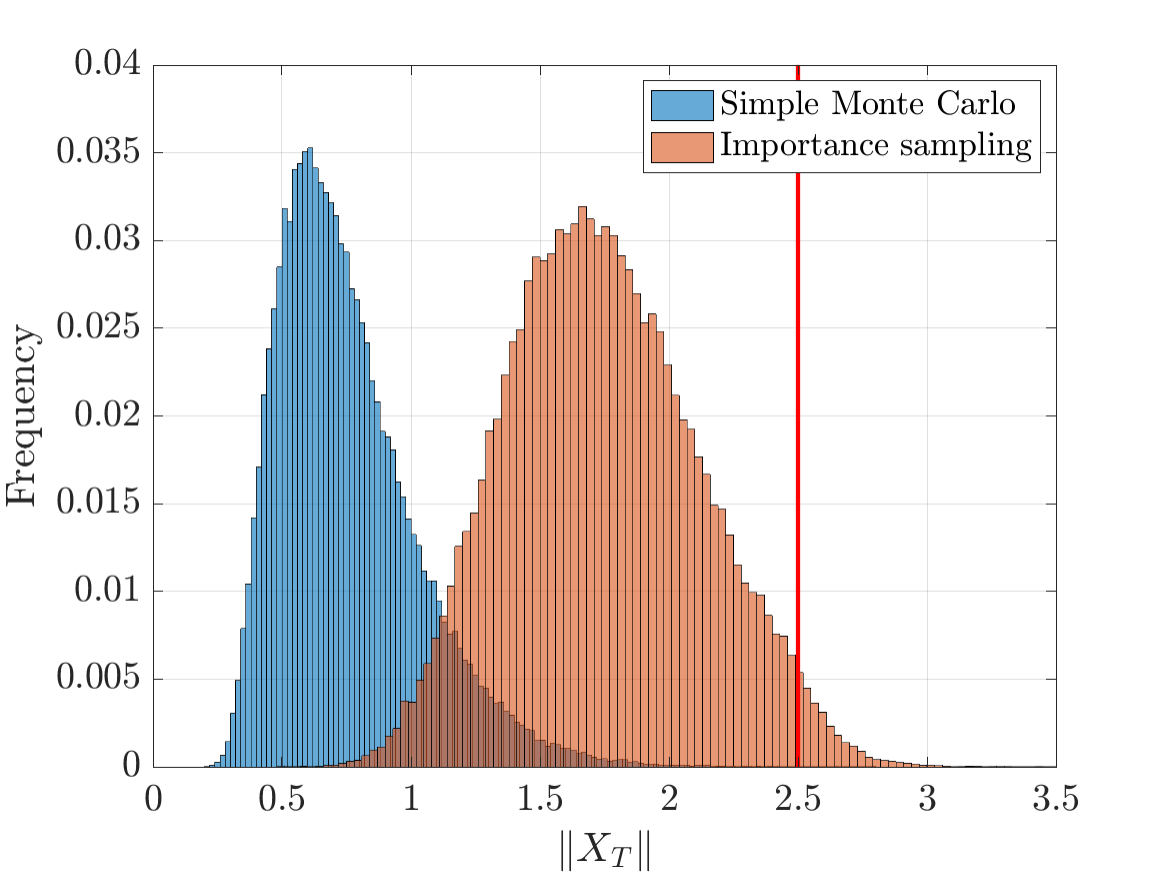}
	\caption{Histograms of $\Vert v(T, \cdot) \Vert_{L^2([0,1])}$ computed using simple Monte Carlo and dynamic importance sampling for the stochastic advection-diffusion equation.}
	\label{fig:advec_distr}
\end{figure}

We plot the histogram of the norm of the system for the biased and unbiased systems in Figure \ref{fig:advec_distr}, and present the results of the sampling methods in Table \ref{tab:advec_diff}. Using only two eigenfunctionals, the variance of the estimator is reduced by a factor of 12 over Monte Carlo.

\begin{table}[H]
\centering
\begin{tabular}{llll}
\hline
\multicolumn{1}{|l|}{}   & \multicolumn{1}{l|}{Variance}             & \multicolumn{2}{l|}{Relative error} \\ \hline
\multicolumn{1}{|l|}{Monte Carlo} & \multicolumn{1}{l|}{$2.02\times 10^{-5}$} & \multicolumn{2}{l|}{$222.5$}     \\ \hline
\multicolumn{1}{|l|}{Importance sampling} & \multicolumn{1}{l|}{$1.68\times10^{-6}$} & \multicolumn{2}{l|}{$64.24$}      \\ \hline
\multicolumn{4}{l}{$\rho_{\text{true}} = 2.02\times10^{-5}$}                                         
\end{tabular}
\caption{Importance sampling performance for the stochastic advection-diffusion equation. }
\label{tab:advec_diff}
\end{table}

\subsection{Nonlinear examples}
\label{subsec:numerical}

\subsubsection{Van der Pol oscillator}
We now demonstrate our approach on nonlinear stochastic systems. Consider the noisy Van der Pol oscillator given by,
\begin{align}
	\de \begin{bmatrix}
		x_1 \\ x_2
	\end{bmatrix} = \begin{bmatrix}
		x_2 \\ m(1-x_1^2)x_2 - x_1 
	\end{bmatrix} \de t + \sqrt{2\epsilon}\begin{bmatrix}
		\de W_1 \\ \de W_2 
	\end{bmatrix}.
\end{align}
In the absence of noise, the system exhibits a limit cycle, such that all initial conditions converge to it (except the origin, which is an unstable equilibrium). In the presence of noise, trajectories cluster on a band that is centered on the limit cycle of the deterministic system. We consider the problem of `peeling' a solution of the stochastic system from this band. Let $m = 0.3$, $\epsilon = 0.01$, and $T = 10$; our task is to estimate
\begin{align}
	\Pb\left[ x_1(T)^2 + x_2(T)^2 >2.7^2 \, \, \vert \, \,  x_1(0) = 2, \,x_2(0) = 0 \right].\label{eq:probvdp}
\end{align}
The initial condition lies on the limit cycle of the deterministic system. The rare event is a region that lies outside of it. 

We first find the sKO eigenfunctions of the system. As described in Section~\ref{sec:dmd}, we apply gEDMD, using a basis $\{\psi_k(x_1, x_2)\}_{k=1}^n$ of bivariate Legendre polynomials with total degree up to 10. This basis is constructed such that it is orthonormal with respect to the uniform measure on $\mathcal{D} = [-4,4]^2 \subset \R^2$. There are $n=66$ elements in this basis. We generate test points by using trajectory data beginning at 400 initial conditions uniformly spaced on $\mathcal{D}$.  Each trajectory is simulated on the interval $t \in [0,10]$. The test points are generated by sampling the trajectories at intervals of $\Delta t = 0.05$, for a total of $8\times 10^4$ test points. 

\begin{figure}[H]
\centering
	\includegraphics[width = 0.5\textwidth]{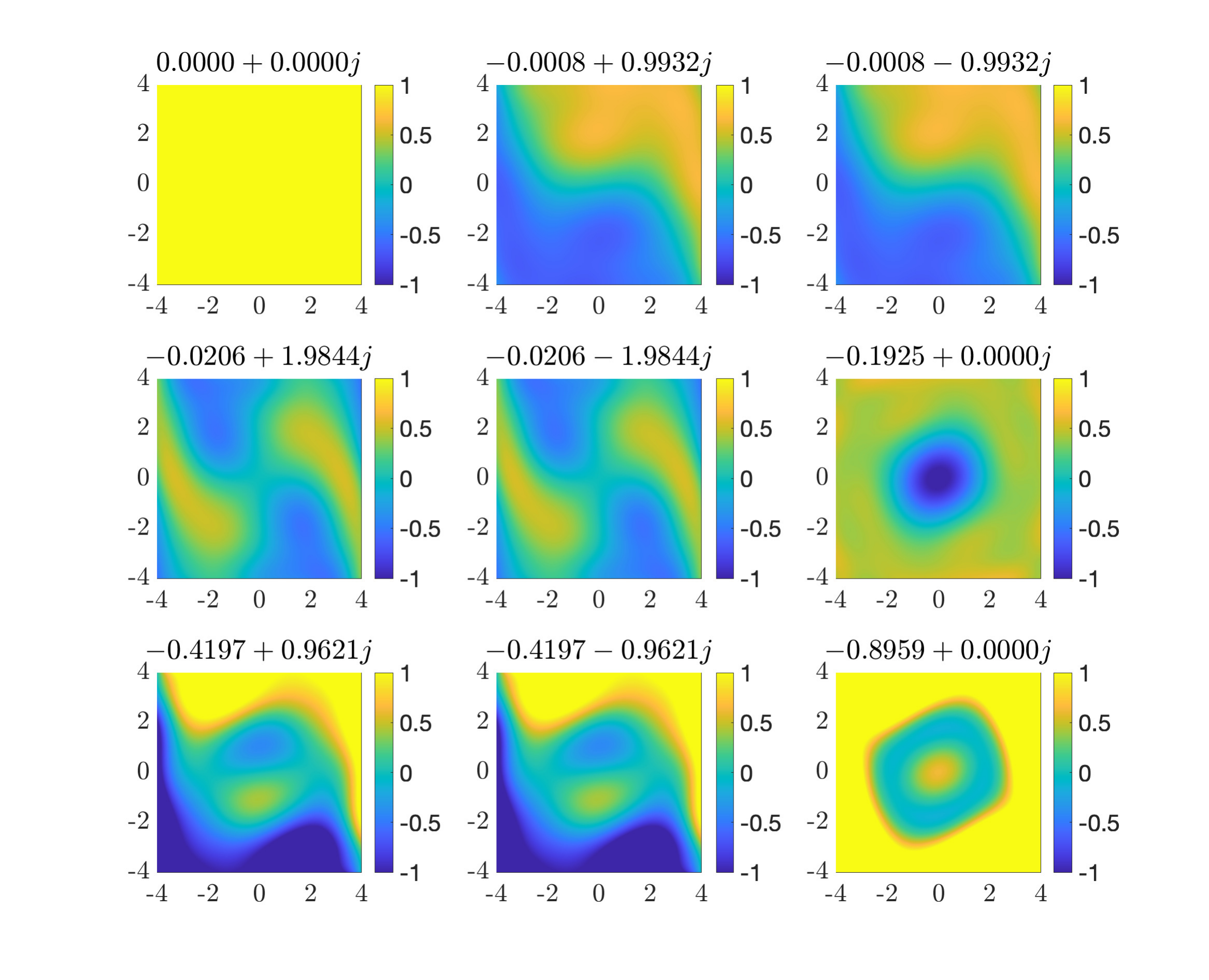}
	\includegraphics[width = 0.49\textwidth]{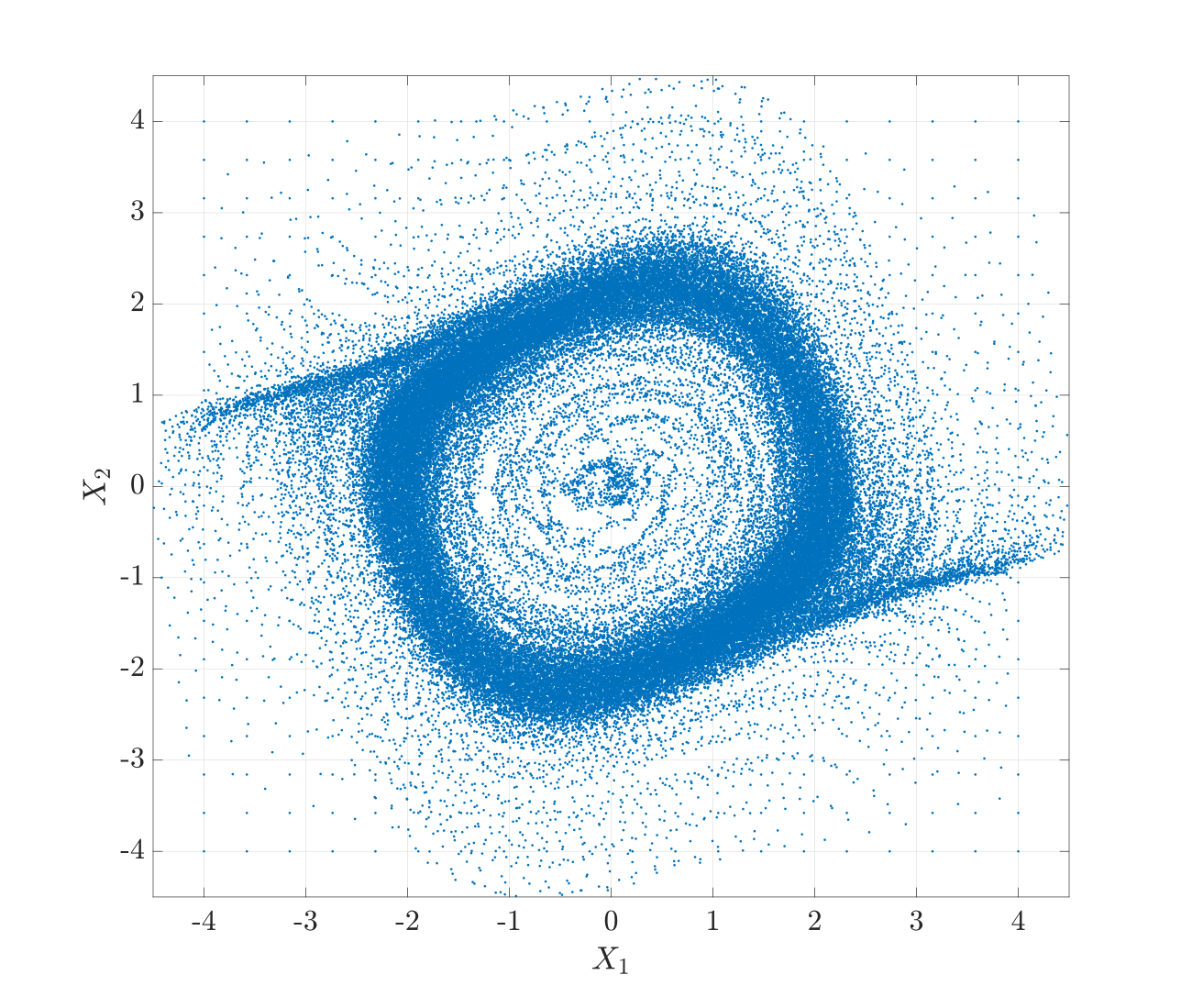}
	\caption{On the left, the first nine stochastic Koopman eigenfunctions for the Van der Pol oscillator. Eigenfunctions are ordered according to the magnitude of the real part of the Koopman eigenvalues, \revi{and only the real part of each eigenfunction is plotted.} Right figure shows the test points.}
	\label{fig:vdpkoop}
\end{figure}

We find that the quality of the eigenvalues and eigenfunctions obtained via gEDMD is highly sensitive to the polynomial degree and the choice of basis. Indeed, it is well-noted that EDMD methods can often lead to spurious eigenvalues, i.e., eigenvalues that are non-physical, when the choice of basis is poor \cite{klus2019datadriven,brunton2019data}. We find that the same is true for \emph{eigenfunctions} obtained via gEDMD, when either the basis is not sufficiently representative of the eigenfunctions or the test points do not sufficiently cover the state space. Obtaining good approximation of the eigenfunctions is critical to our sampling framework. Given a set of candidate eigenfunctions produced by gEDMD, we cross-validate them with an independent dataset generated in the same fashion as the test points. In particular, the mean-square error of a candidate eigenfunction $\phi(x)$ with eigenvalue $\lambda$ is defined as $\frac{1}{m} \sum_{i = 1}^m |\mathcal{A}\phi(x_i) - \lambda \phi(x_i) |^2$. Only candidate eigenfunctions with a testing error below some threshold (here chosen to be 0.04) are used to approximate the Doob transform. In Figure \ref{fig:vdpkoop} we show the first nine approximated \revi{(and validated)} sKO eigenfunctions, alongside a scatterplot of the test points. 

Approximating the Doob transform to estimate \eqref{eq:probvdp} requires approximating the indicator function over the rare event region in the sKO eigenbasis. We first express the indicator function in the Legendre basis by solving a least-squares problem on the gEDMD test points. Since the Koopman eigenfunctions are approximated in the same basis, we can immediately compute the coefficients of the indicator's sKO eigenfunction expansion. 
Just as in the linear case, if the expansion in the sKO eigenfunction basis is negative in some region of the domain of interest, we add a constant \revi{so that the approximation to $f(x)$ has value greater than $0.01$. A scaling factor of $c = 17$ is again applied to the biasing so that enough samples will reach the rare event.} In this example, we use fifteen eigenfunctions, of which nine are plotted in Figure \ref{fig:vdpkoop}, to approximate the Doob transform. 

\begin{figure}[H]
\centering
	\includegraphics[width = 0.49\textwidth]{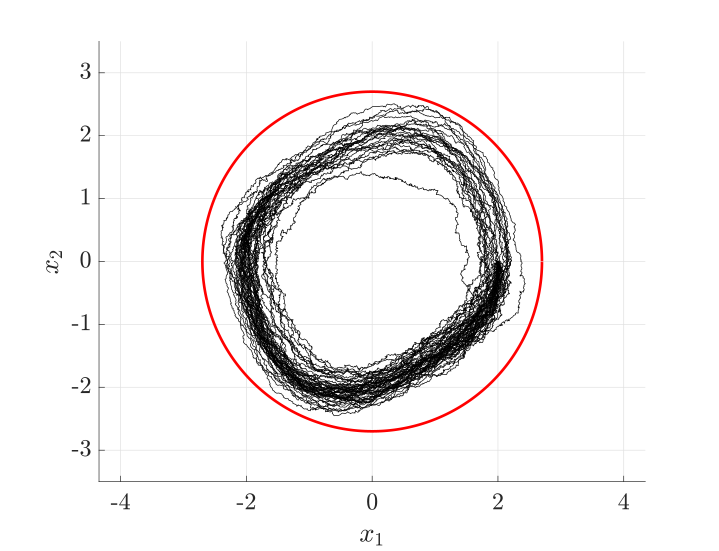}
	\includegraphics[width = 0.49\textwidth]{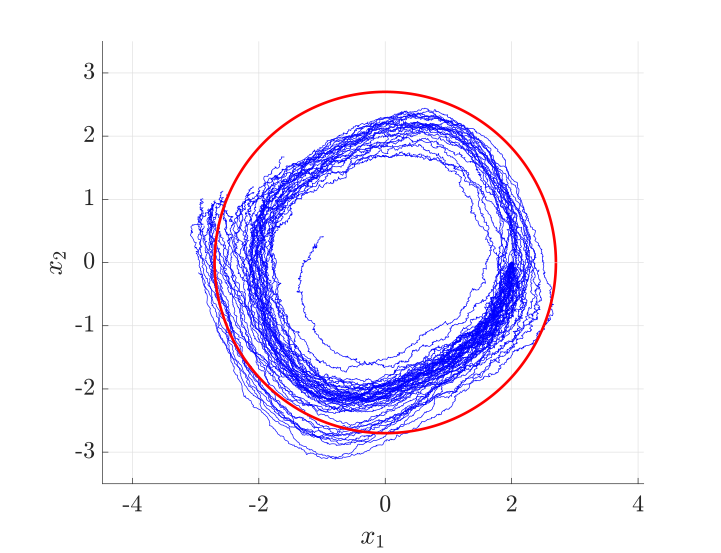}
	\caption{Left: sample paths of unbiased Van der Pol oscillator. Right: sample paths of biased Van der Pol oscillator. Red circle denotes boundary of the rare event. }
	\label{fig:vdppaths}
\end{figure}

\begin{figure}[H]
	\centering
	\includegraphics[width = 0.5\textwidth]{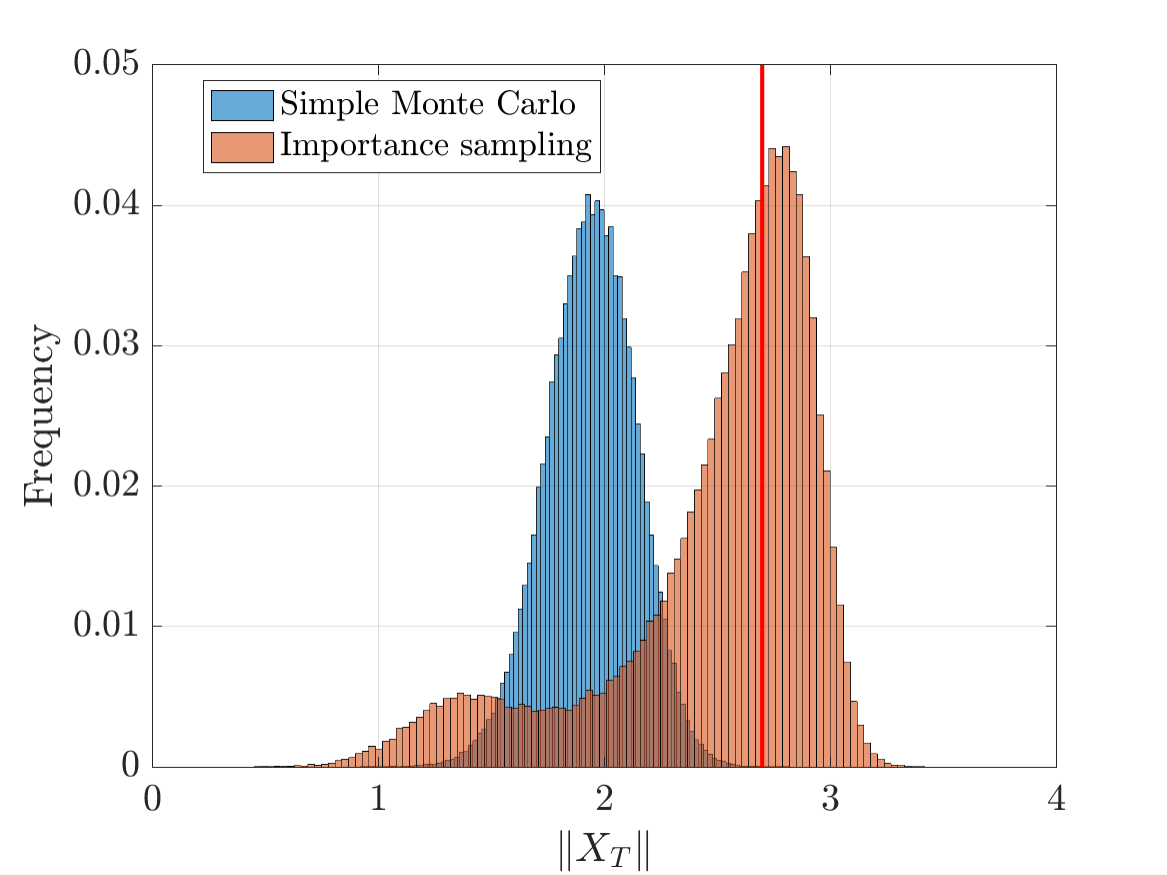}
	\caption{Distribution of norm of Van der Pol state at time $T = 10$. Red line denotes boundary of the rare event. }
	\label{fig:vdp_distr}
\end{figure}

In Figure \ref{fig:vdppaths}, we show 25 unbiased and biased sample paths of the oscillator. Notice that none of these unbiased sample paths reaches the rare event---they all remain inside the red circle demarcating the rare event region---while many of the biased paths do reach it. In Figure \ref{fig:vdp_distr}, we show the histogram of norm of the state at time $T = 10$ for the two systems. We report simulation results for the estimators in Table \ref{tab:vdp}, and observe that the importance sampling estimator reduces variance by a factor of more than 400. 

\begin{table}[H]
\centering
\begin{tabular}{llll}
\hline
\multicolumn{1}{|l|}{}   & \multicolumn{1}{l|}{Variance}             & \multicolumn{2}{l|}{Relative error} \\ \hline
\multicolumn{1}{|l|}{Monte Carlo} & \multicolumn{1}{l|}{$1.69\times 10^{-5}$} & \multicolumn{2}{l|}{$243.01$}     \\ \hline
\multicolumn{1}{|l|}{Importance sampling} & \multicolumn{1}{l|}{$4.03\times10^{-8}$} & \multicolumn{2}{l|}{$11.85$}      \\ \hline
\multicolumn{4}{l}{$\rho_{\text{true}} = 1.69\times10^{-5}$}                                            
\end{tabular}
\caption{Importance sampling performance for the van der Pol oscillator.}
\label{tab:vdp}
\end{table}

\subsubsection{Duffing oscillator}
Now we consider the noisy Duffing oscillator,
\begin{align*}
	\ddot{x} + \delta \dot{x} + x(\beta + \alpha x^2) = \sqrt{2\epsilon} \,\dot{W}_t,
\end{align*}
which can be rewritten in standardized form as
\begin{align}
	\de \begin{bmatrix}
		x_1 \\ x_2
	\end{bmatrix} = \begin{bmatrix}
		x_2 \\ -\delta x_2 - x_1(\beta+\alpha x_1^2)
	\end{bmatrix} \de t + \sqrt{2\epsilon}\begin{bmatrix}
		0\\ \de W_t 
	\end{bmatrix}.
\end{align}
The deterministic Duffing oscillator has three equilibria. The origin is an unstable equilibrium, while $x^{*} = \pm \sqrt{-\beta/\alpha}$ are two stable equilibria. In the basins of attraction of the stable equilibria, the system exhibits damped oscillatory dynamics.  In the stochastic setting, noise can infrequently cause trajectories to transition between the basins of attraction. For a transition to occur, the stochastic forcing must ``kick'' the system in the correct direction and with the correct magnitude, in critical regions of the state space. We thus consider the rare event of \emph{transitioning} from one basin of attraction to the other: 
\begin{align*}
	\Pb\left[x_1(T)>0 \, | \, x_1(0) = -1.5, \, x_2(0) = 0 \right]. \label{eq:dduffingprob}
\end{align*}
Here we use parameter values $\alpha = 1$, $\beta = -1$, $\delta = 0.5$, $\epsilon = 0.0025$, and $T = 10$. The study of noise-induced transitions between attractors in dynamical systems is an important problem that arises in protein folding and chemical kinetics \cite{vanden2006transition,vanden2012rare,chiavazzo2017intrinsic}. 

\begin{figure}
	\centering
	\includegraphics[width = 0.5\textwidth]{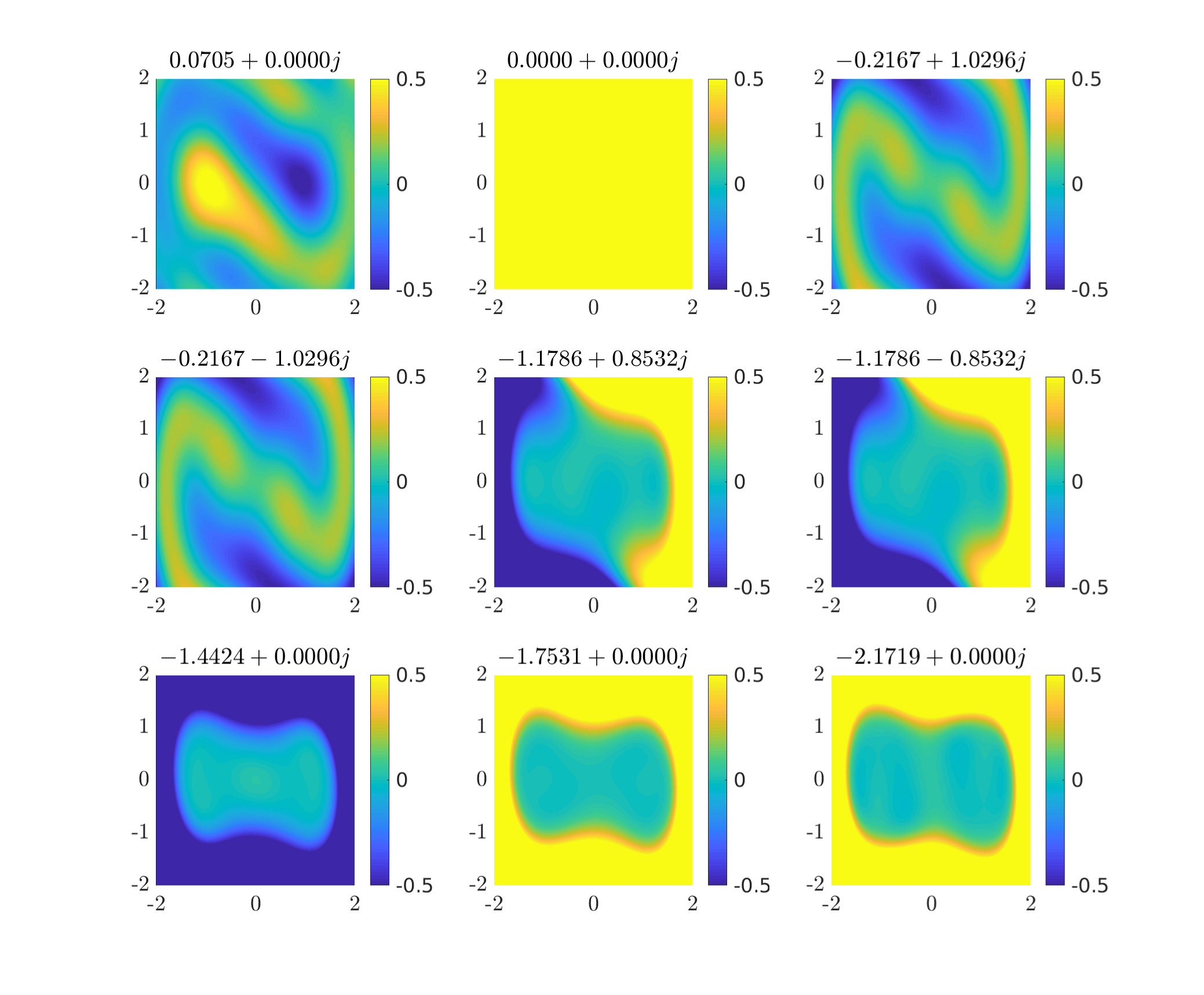}
	\includegraphics[width = 0.49\textwidth]{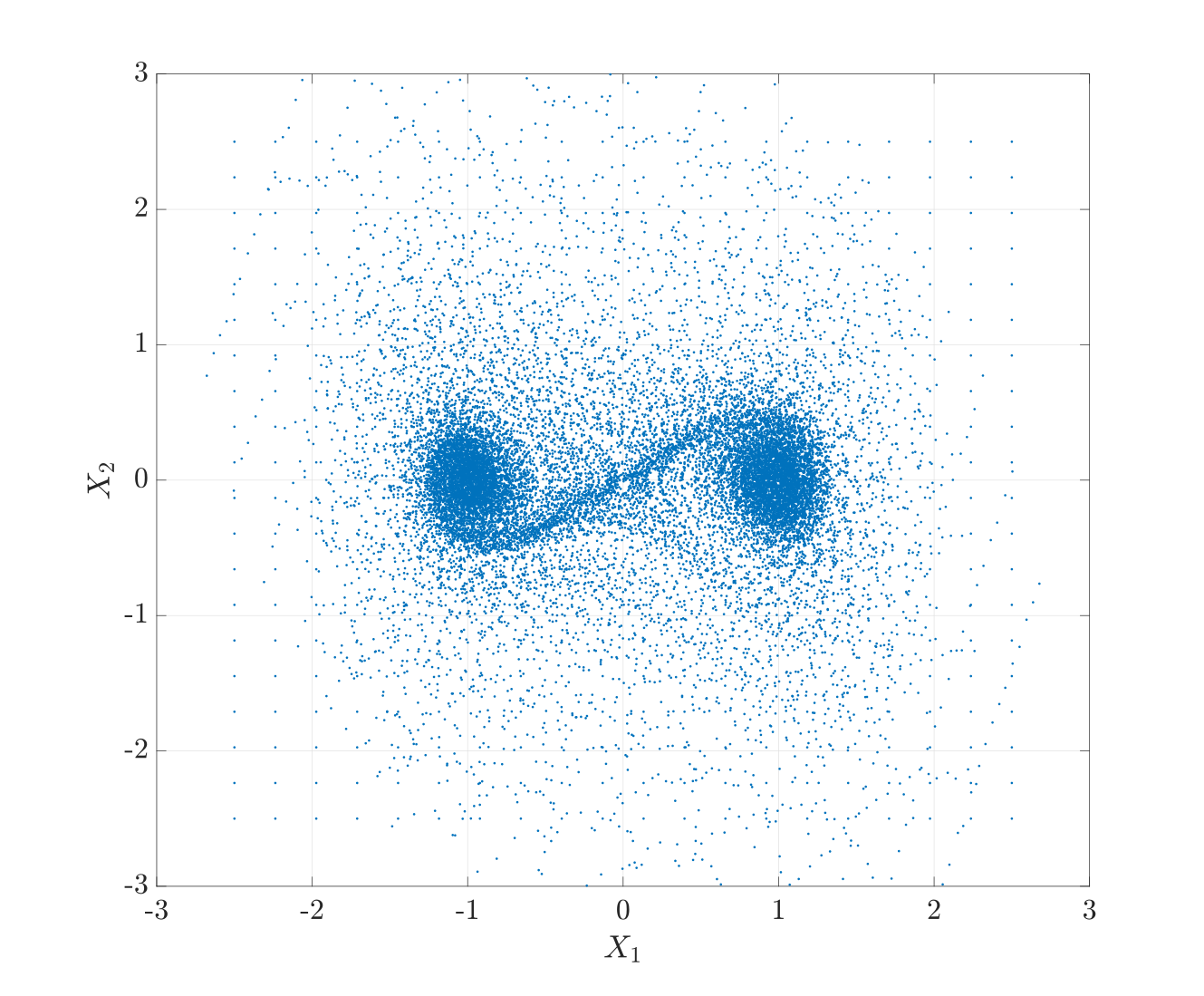}
	\caption{First nine stochastic Koopman eigenfunctions of the noisy Duffing oscillator. Eigenfunctions are ordered according to the magnitude of the real parts of the Koopman eigenvalues.}
	\label{fig:duffingeigfunc}
\end{figure}

Similar to the Van der Pol oscillator, we find the stochastic Koopman eigenfunctions by applying gEDMD with a basis of bivariate scaled Legendre polynomials of total degree up to 12. This leads to 91 basis functions. To create test points for gEDMD, we simulate 400 independent trajectories over the interval $t \in [0,10]$, with initial conditions uniformly spaced over $\mathcal{D} = [-2.5,2.5]^2$. The data set is then generated by sampling each trajectory at intervals of $\Delta t = 0.2$. In Figure \ref{fig:duffingeigfunc}, we show the first nine \revi{approximated and validated} eigenfunctions of the stochastic Duffing oscillator, along with the scatterplot of the test points.
We approximate the indicator function $f(x) = \mathbbm{1}_{x_1 > 0}(x_1, x_2)$ via a linear combination of these nine sKO eigenfunctions, using regression on the same test points. \revi{As before, we add a constant to the approximation so that the minimum value of the approximation to $f(x)$ is greater than $0.01$, and scale the biasing term with a multiplicative factor $c = 8$.}

\begin{figure}
	\centering
	\includegraphics[width = 0.49\textwidth]{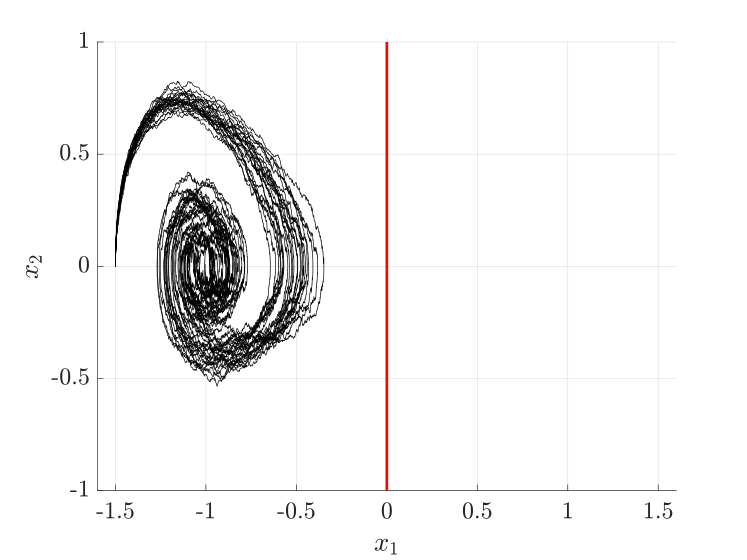}
	\includegraphics[width = 0.49\textwidth]{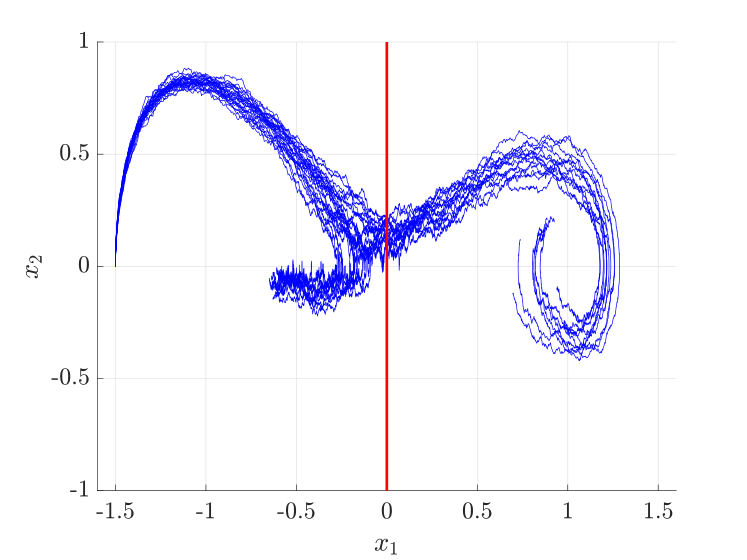}
	\caption{Left: sample paths of the unbiased Duffing oscillator. Right: sample paths of the biased Duffing oscillator. Red line denotes the boundary of the rare event.}
	\label{fig:duffingpaths}
\end{figure}

\begin{figure}
\centering
	\includegraphics[width = 0.5\textwidth]{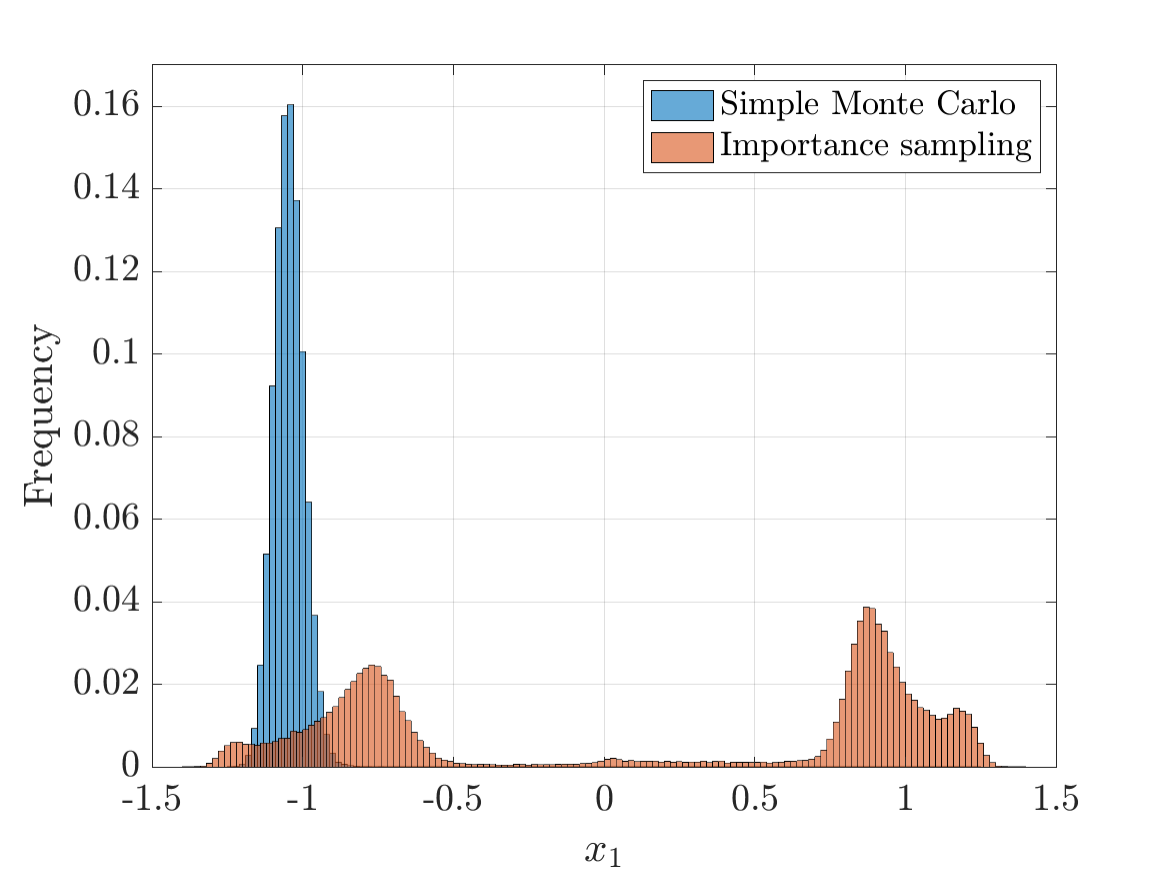}
	\caption{Noisy Duffing oscillator: histogram of $x_1$ at time $T=10$ for the unbiased and biased systems.}
	\label{fig:duffing_distr}
\end{figure}

In Figure \ref{fig:duffingpaths}, we show 25 of the resulting biased sample trajectories of the Duffing oscillator, compared to unbiased paths. The few unbiased trajectories shown here do not transition to the opposite basin of attraction. We plot a histogram of the final positions of the unbiased and biased sample trajectories in Figure \ref{fig:duffing_distr}. The figure demonstrates that unlike simple Monte Carlo, the biased trajectories are able to sample the transition paths with much greater success. Quantitative performance of the estimators is compared in Table \ref{tab:Duffingrestults}.  In particular, it can be seen that the importance sampling estimator reduces  variance by a factor of nearly 5000. 

\begin{table}
\centering
\begin{tabular}{llll}
\hline
\multicolumn{1}{|l|}{}   & \multicolumn{1}{l|}{Variance}             & \multicolumn{2}{l|}{Relative error} \\ \hline
\multicolumn{1}{|l|}{Monte Carlo} & \multicolumn{1}{l|}{$2.11\times 10^{-5}$} & \multicolumn{2}{l|}{$217.93$}     \\ \hline
\multicolumn{1}{|l|}{Importance sampling} & \multicolumn{1}{l|}{$4.35\times10^{-9}$} & \multicolumn{2}{l|}{$3.13$}      \\ \hline
\multicolumn{4}{l}{$\rho_{\text{true}} = 2.11\times10^{-5}$}                                            
\end{tabular}
\caption{Importance sampling performance for the noisy Duffing oscillator.}
\label{tab:Duffingrestults} 
\end{table}

\section{Analyzing the second moment}
\label{sec:analysis}

It is useful to understand how the approximation of the Doob transform impacts the variance of the resulting importance sampling estimator. Here we provide some simple analytical results to that end. For this analysis, we assume that the sKO eigenfunctions are obtained exactly. Consequently, the only error in the solution to the KBE originates from the accuracy of approximation of the terminal condition. 

We perform a non-asymptotic analysis of the importance sampling scheme based on the approach outlined in \cite{dupuis2015escaping,dupuis2012importance}. Assume $f(x) \ge 0$ and define $h(x) = -\log f(x)$; recall that $f(x)$ is the terminal condition of the KBE in \eqref{eq:kbe}, which is typically the indicator function over the rare event. In contrast to our earlier presentation of the Doob transform, here $f$ is allowed to be a true indicator function, rather than a mollified version of it. Indeed, the analysis in \cite{dupuis2015escaping,dupuis2012importance} takes this scenario into account. Recall that the importance sampling estimator (cf.\ \eqref{eq:isest}) of $\rho = \Ex^{0,x_0}[f(X_T)]$ can be written as 
\begin{align*}
	\Gamma(x_0) = e^{-h(\tilde{X}_T)} \frac{\de \Pb}{\de \mathbb{Q}}(\tilde{X}),
\end{align*} 
where $x_0$ is the initial condition. The second moment of the importance sampling estimator corresponding to any control $u(t,x)$ in the SDE system \eqref{eq:sdealt} is 
\begin{align*}
	Q(x_0;u) = \Ex_{{\mathbb{Q}}} \left[ e^{-2h(\tilde{X}_T)}\left(\frac{\de \Pb}{\de \mathbb{Q}} \right)^2\right]. 
\end{align*}
Using \cite{boue1998variational} and the subsequent analysis in \cite{dupuis2015escaping,dupuis2012importance}, we obtain the following variational representation of the second moment of the importance sampling estimator: 
\begin{align}
	-\log Q(x_0; u) = \inf_{v\in\mathcal{V}} \Ex\left[\frac{1}{2}\int_0^T \|v(s)\|^2 \,\de s - \int_0^T \|u(s,\hat{X}_s)\|^2 \, \de s + 2 h(\hat{X}_T)\right],
\end{align}
where $\hat{X}$ solves
\begin{align*}
\begin{dcases}
		\de \hat{X}_s = \left[\A(\hat{X}_s) - \B(\hat{X}_s)u(s,\hat{X}_s) + \B(\hat{X}_s) v(s) \right] \, \de s + \B(\hat{X}_s) \de W_s \\
		\hat{X}_0 = x_0
\end{dcases}\, ,
\end{align*}
and $\mathcal{V}$ is the set of progressively measurable admissible processes. Recall that when $f$ is the indicator function over set $E$, $h(\hat{X}_T)$ is infinity if $\hat{X}_T$ does not enter $E$ and zero otherwise. We can, therefore, restrict the set of admissible processes so that $\mathcal{V}$ only contains controls ensuring $\hat{X}_T\in E$ with probability one. Then by Lemma A.1 in \cite{dupuis2015escaping}, we have that for any sufficiently regular functions $Z(t,x)$ and $U(t,x)$, where the control is $u = - \B^*\nabla U$, 
\begin{align}
	-\log Q(x_0;u) \ge \inf_{v\in\mathcal{V}} 2 U(0,x_0)& - 2\Ex[U(T,\hat{X})] + 2 \int_0^T \mathcal{G}[Z](s,\hat{X}) \, \de s \label{eq:bound} \\ &- \int_0^T \|\mathbf{B}^*(\nabla Z - \nabla U) \|^2 \, \de s  \, , \nonumber
\end{align}
where $\mathcal{G}[Z] = \partial_t Z + \langle \mathbf{A}(x), \nabla Z \rangle - \frac{1}{2}\|\mathbf{B}^* \nabla Z \|^2 +\frac{1}{2} \text{Tr} \mathbf{BB}^* \nabla^2 Z.$ The operator $\mathcal{G}$ can be obtained from the partial differential operator of the KBE, $\partial_t [\cdot] + \mathcal{A}[\cdot]$, via a change of variables $Z = -\log \Phi $.  

In our approach, the controller is derived from an approximation to the solution of the KBE: $\tilde{u}(t,x) = \mathbf{B}(x)\nabla \log \tilde{\Phi}(t,x)$. Therefore, if we choose $U(t,x) = -\log \tilde{\Phi}(t,x)$, then we can use \eqref{eq:bound}. Recall that we have assumed $\tilde{\Phi}$ \eqref{eq:approxKBEsol} to be constructed with the exact sKO eigenfunctions. Therefore, it is an exact solution of the KBE of the system for $t \in [0,T)$, but does not match the terminal condition at $t = T$. Nonetheless, we have $\mathcal{G}[U] = 0$ exactly. Taking $Z=U$ then gives
\begin{align}
	-\log Q(x_0;u) \ge 2 U(0,x_0) -2 \sup_{v\in\mathcal{V}} \Ex\left[ U(T,\hat{X}_T) \right]. 
	\label{eq:bound2}
\end{align}
Note that the above bound is tight if $\tilde{\Phi}(t,x)$ in fact exactly matches the terminal condition, $\tilde{\Phi}(T,x) = f(x)$. In this case, we have that $U(T,\hat{X}_T) = h(\hat{X}_T)$, which, in turn, implies that the right-hand side equals $2U(t,x) = -2\log \rho$ and therefore $Q(x_0;u) \le \rho^2$. Since $Q(x_0;u)\ge \rho^2$ by Jensen's inequality, we conclude $Q(x_0;u) = \rho^2$. In other words, the variance of the estimator is zero. 

On the other hand, when the biasing is imperfect but based on the true sKO eigenfunctions, \eqref{eq:bound2} implies that the bound on the second moment depends on the accuracy of the approximations of $f(x)$ and of the KBE solution at the initial condition $x_{0}$ (i.e., the quantity of interest $\rho = \Phi(0,x_0)$) using the eigenfunctions. Recalling that $U(T,\hat{X}_T) = -\log \tilde{\Phi}(T,\hat{X}_T)$, observe that
\begin{align*}
	-2\sup_{v\in \mathcal{V}} \Ex\left[ U(T,\hat{X}_T)\right] = 2\inf_{v\in\mathcal{V}} \Ex\left[ \log \tilde{\Phi}(T,\hat{X}_T)\right]. 
\end{align*}
Appealing to the properties of the expectation and the fact that $\hat{X}_T\in E$ with probability one, 
\begin{align*}
	\Ex\left[\log \tilde{\Phi}(T,\hat{X}_T) \right]\ge \inf_{y\in E} \Ex\left[\log \tilde{\Phi}(T,y) \right] = \inf_{y \in E} \log \tilde{\Phi}(T,y).
\end{align*}
Then \eqref{eq:bound2} can be bounded from below as follows: 
\begin{align}
	-\log Q(x_0;u) \ge -2 \log \tilde{\Phi}(0,x_0) + 2\inf_{y \in E} \log \tilde{\Phi}(T,y).
	\label{eq:boundsecmom}
\end{align}
This relation is an upper bound for the second moment for the importance sampling estimator. The first term reflects how well the solution approximates the quantity of interest $\rho$. The second term reflects how well the approximate KBE solution approximates the terminal condition in the rare event. It is important to emphasize that these two terms are coupled to one another since the solution at the initial condition is dependent on how well the terminal condition is approximated. 

Further refinement of these bounds is difficult. The framework we have presented is rather general, in the sense that we did not make strong assumptions on the properties of the stochastic dynamical system. Moreover, without prescribing closed-form or otherwise very specific approaches to positivization or scaling (i.e., choosing $\varepsilon$ and $c$), it is difficult to characterize precisely how well the sKO eigenfunctions approximate the solution to the KBE. For specific classes of dynamical systems, one might be able to elucidate these bounds further, but since the emphasis of this paper has been on a generally applicable computational approach, we leave such analyses to future work. 

\section{Conclusions}
\label{sec:discussion}

We have presented a framework for constructing importance sampling schemes for stochastic dynamical systems, using eigenfunctions of the associated stochastic Koopman operator (sKO). We use sKO eigenfunctions to approximate the Doob transform for the observable of interest, which in turn yields an approximation of the corresponding zero-variance importance sampling estimator. Our approach is broadly applicable, and we demonstrate the computation of rare event probabilities in a wide variety of linear and nonlinear SDEs. 
These numerical examples highlight how one can exploit non-rare (bulk) trajectories of the dynamical system to inform biasing strategies for rare event simulation. For systems where the sKO eigenfunctions cannot be derived analytically, we used generator EDMD to \revi{approximate} them numerically. Our approach is agnostic to the numerical method used to approximate the sKO eigenfunctions, however, and thus as state-of-the-art methods for numerical approximation of the Koopman operator improve, our framework too will improve in accuracy and efficacy. Moreover, even imprecise applications of our approach can still lead to significant variance reduction. We demonstrate that crude approximations to the Doob transform, using only a few numerically-approximated eigenfunctions, can lead to variance reduction of several orders of magnitude over simple Monte Carlo.

We note that our approach is applicable to a wide range of stochastic dynamical systems, including many that are not typically handled by existing rare event simulation methods. Methods inspired by computational chemistry, for example, typically consider high-dimensional diffusion processes governed by a potential, i.e., gradient systems. We instead propose a {single} framework that enables rare event simulation in systems with non-normal dynamics, oscillatory behavior, limit cycles, and degenerate noise---which appear in a variety of scientific and engineering settings \cite{cousins2019predicting,zhang2018rare, strogatz2018nonlinear}. This framework often ``recovers'' solutions proposed for specific cases. For instance, in non-normal linear systems, we find that the dominant direction of biasing away from an attracting point is aligned with the leading left eigenvectors of the drift term. This is consistent with the rigorous theoretical results of \cite{salins2016rare} for infinite-dimensional \emph{self-adjoint} linear systems, where the left and right eigendirections coincide; there, the authors found that (given a sufficient spectral gap) the best way to escape from an attractor is again to bias in the direction of the most slowly decaying eigenmode.

We can also contrast our approach with rare event simulation methods based on stochastic optimal control \cite{hartmann2017variational,hartmann2019variational,zhang2014applications}. The goal of these efforts is the same as ours: to find a controller for the dynamical system that approximates the zero-variance importance sampling estimator. However, the stochastic optimal control formulation requires solving optimization problems or the associated nonlinear Hamilton-Jacobi-Bellman equation, both of which may be intractable in high dimensions. These methods attempt to \emph{precisely} compute the Doob transform \emph{locally}, depending on where trajectories lie in state space. In contrast, we consider the Kolmogorov backward equation, which, due to linearity, enables efficient computation based on eigenfunction information. Our approach thus \emph{crudely} computes the Doob transform \emph{globally}, using sKO eigenfunctions approximated via non-rare trajectories. 

There are several avenues for future work. For example, approximation of the terminal condition of the KBE via sKO eigenfunctions presents some outstanding questions. We currently construct this approximation by combining regression with a post hoc numerical correction to ensure positivity. A single integrated, consistent procedure for constructing positive approximations would be preferable: not only might it improve the efficiency of rare event simulation, but it could also enable further theoretical analysis of approximation error and hence estimator variance.

\revi{ 
A practical bottleneck of our framework is the accuracy to which DMD methods can approximate the sKO eigenfunctions in regimes where the amount of data is limited relative to the dimensionality of the problem. Addressing this issue will be useful for scaling our approach to more complex high-dimensional systems. The resolution depends in part on how Koopman numerical methods develop in the future. Our current approach requires the ability to evaluate gradients of eigenfunctions anywhere in the state space. State-of-the-art Koopman numerical methods for high-dimensional systems such as Hankel DMD only give values of the eigenfunctions at the test points. The key to addressing these problems will be to find an alternative to importance sampling that uses only the eigenfunctions, and not their gradients, for sampling.} Moreover, importance sampling is known to be, in many cases, an unstable method, as there might be no guarantees that the variance of the resulting estimator is finite \cite{budhiraja2019analysis}. In future work, we plan to adapt our approach to more robust sampling methods for rare event simulation, such as multilevel splitting. The connection between efficient importance sampling estimators and multilevel splitting has been well established  \cite{budhiraja2019analysis}. Multilevel splitting also has the virtue of being non-intrusive, meaning that one is not required to alter the system dynamics to perform rare event simulation. Since Koopman numerical methods enable us to construct crude approximations to the KBE non-intrusively, combining these methods with multilevel splitting will lead to more efficient \emph{black box} approaches for rare event simulation. We hope that these avenues for future research will produce new algorithms for robust, data-driven rare event simulation in diverse applications.

\section*{Acknowledgements}
This material is based upon work supported by the DARPA EQUiPS program through the United States Air Force under contract number FA8650-16C-7646 and by the Air Force Office of Scientific Research under contract number FA9550-20-1-0397. We would also like to thank Konstantinos Spiliopolous, Paul Dupuis, and Jose Blanchet for their helpful advice and insights. 
\appendix

\section{Proof of Theorem 1}
\label{app:doobh}
\revi{While the Doob $h$-transform is a standard result \cite{rogers2000diffusions,sarkka2019applied}, we provide a proof of the particular form presented in Theorem \ref{theo:doob} to elucidate how it is important in the construction of our importance sampling estimator.}

\begin{proof}
	We compute the stochastic integral 
	\begin{align*}
		\int_0^T \langle u(t,\tilde{X}_t),\de W_t\rangle.
	\end{align*}
	Let $g(t,x) = \log \Phi(t,x)$, and apply It\^o's formula: 
	\begin{align*}
		\de g =& \ddt \log \Phi(t,\tilde{X}_t) \de t+ \langle \nabla\log \Phi(t,\tilde{X}_t),\de\tilde{X}_t \rangle  + \frac{1}{2} \text{Tr}\left[\nabla^2[\log \Phi(t,x)] (\de \tilde{X}_t)(\de \tilde{X}_t)^* \right] \\
		=& \frac{1}{\Phi}\ddt \Phi \,\de t + \left\langle \frac{1}{\Phi}\nabla \Phi, \textbf{A}(\tilde{X}_t)+\textbf{BB}^*\frac{\nabla \Phi}{\Phi} \right\rangle \de t + \left\langle \frac{\nabla \Phi}{\Phi},\textbf{B}\, \de W_t \right\rangle  \\
		&+ \frac{1}{2}\text{Tr}\left[\textbf{BB}^* \frac{\nabla^2 \Phi}{\Phi} \right] \de t -\frac{1}{2}\text{Tr}\left[ \textbf{BB}^* \frac{(\nabla \Phi)(\nabla \Phi)^*}{\Phi^2} \right] \de t \\
		=& \frac{1}{\Phi}\left(\ddt \Phi + \langle \nabla \Phi, \textbf{A}(\tilde{X}_t)\rangle + \frac{1}{2}\text{Tr} \left[ \textbf{BB}^* \nabla^2 \Phi\right] \right) \de t+ \frac{1}{2}\left\langle \frac{\textbf{B}^*\nabla \Phi}{\Phi},\frac{\textbf{B}^*\nabla \Phi}{\Phi} \right\rangle \de t \\
		& + \left\langle \frac{\nabla \Phi}{\Phi},\textbf{B}\, \de W_t \right\rangle \\
		=& \frac{1}{\Phi}\left(\frac{\partial\Phi}{\partial t} + \mathcal{A}\Phi \right)\, \de t + \frac{1}{2} \| \textbf{B}^* \nabla \log \Phi(t,\tilde{X}_t) \|^2\, \de t  + \langle \textbf{B}^* \nabla\log \Phi(t,\tilde{X}_t), \de W_t \rangle\\
		=& \frac{1}{2} \|u(t,\tilde{X}_t)\|^2 \de t + \langle u(t,\tilde{X}_t),\de W_t\rangle. 
	\end{align*}
	This implies that
	\begin{align*}
		\int_0^T \langle  u(t,\tilde{X}_t), dW_t\rangle  = \log \Phi(T,x)- \log \Phi(0,x) -\frac{1}{2} \int_0^T \|u(t,\tilde{X}_t)\|^2  \de t.
	\end{align*}
	Plugging this into \eqref{eq:zerovarest}, we have
	\begin{align*}
		f(\tilde{X}_T)\exp\left[-\int_0^T \langle u(t,\tilde{X}_t),\de W_t\rangle - \frac{1}{2}\int_0^T \|u(s,\tilde{X}_s)\|^2 \de s \right] &= f(\tilde{X}_T) \exp\left[ \log \Phi(0,x)- \log \Phi(T,\tilde{X}_T)\right] \\
		& = f(\tilde{X}_T)\frac{\Phi(0,x)}{f(\tilde{X}_T)} \\
		& = \Phi(0,x).
 	\end{align*}
 	Thus, the biasing leads to a zero variance estimator.
\end{proof}

\section{Numerical solution to stochastic PDEs}

To make this paper more self-contained, we provide a brief review of methods for simulating stochastic PDEs. Much of this presentation is based on \cite{jentzen2009numerical,zhang2017numerical}. We use these methods when discretizing the stochastic advection-diffusion equation in Section~\ref{sec:advdiff}. 
 
Stochastic PDEs are typically solved by formulating the equation as a stochastic differential equation on a Hilbert space of functions defined over some subset of $\R^d$. Let $H^2(D)$ be a Sobolev space over an open set $D \subset \R^d$. Let $\bfA$ be a compact self-adjoint linear operator that maps $H^2(D)$ to itself, and $f$ be a possibly nonlinear function from $H^2(D)$ to itself. Stochastic PDEs are typically formulated in the following semilinear form,
\begin{align}
	 \de X_t = \left[\textbf{A}X_t + f(X_t) \right] \, \de t + \de W_t,
\end{align}
where $f$ are nonlinear functions from $H^2(D)$ to itself, $X_t \in H^2(D)$ for all $t$, and $W_t$ is an infinite-dimensional Wiener process. The inner product over this Sobolev space is
\begin{align}
	\langle \psi,\phi\rangle = \int_D \psi \phi\, \de x.
\end{align}
Theoretical details on infinite-dimensional Wiener processes can be found in \cite{da2014stochastic}.

\subsection{Simulating infinite dimensional Wiener processes}
Let $H$ be a separable infinite dimensional Hilbert space and $W_t$ be an $H$-valued $Q$-Wiener process. One may simulate this process using a series expansion. Let $\{e_k\}_{k \in \mathbb{N}}$ be an orthonormal basis of $H$ comprised of eigenvectors of $Q$ with eigenvalues $q_k>0$. One can then represent $W_t$ as follows,
\begin{align}
  	W_t = \sum_{k = 1}^\infty \sqrt{q_k} \beta_k (t) e_k,
  \end{align}  
  where $\beta_k$ are independent real-valued one-dimensional Wiener processes.

\subsection{Exponential Euler schemes}\label{sec:euler}
The exponential Euler scheme is a type of Galerkin method in which the linear part of the projected SPDE is solved exactly. The nonlinear parts are integrated forwards in time using Duhamel's principle. We assume $\bfA$ admits an orthonormal basis $\{\phi_i\}_{i = 1}^\infty$ in $L^2(D)$ with eigenvalues $-\lambda_i$ for $\lambda_i>0$, where $\phi_k \in H^2(D) \cap H_0^1(D)$. Here, $H_0^1(D)$ denotes the Sobolev space whose functions satisfy Dirichlet boundary conditions. 

Define a finite-dimensional subspace of $H^2(D)$ via a subset of the orthonormal basis $\{\phi_i\}_{i = 1}^N$. We project the SPDE onto the resulting $N$-dimensional space and obtain a finite-dimensional representation of the SPDE,
\begin{align}
	\de X_t^N = \left[ \bfA_N X_t^N + F_N(X_t^N) \right] \, \de t +  \de W_t^N
\end{align}
where
\begin{align}
	\textbf{A}_N v = \sum_{i = 1}^N -\lambda_i \langle v,\phi_i\rangle \, \phi_i \, , \\
	F_N = \mathcal{P}_N F |_{\mathcal{X}_N} = \sum_{i = 1}^N \langle f(X_t^N),\phi_i\rangle \, \phi_i.
\end{align}
The result of the projection is called the It\^o-Galerkin stochastic ODE. The mild representation of the solution is
\begin{align}
	X_t^N = e^{\bfA_N t} x_0^N + \int_0^t e^{\bfA_N(t-s)} F_N(X_s^N) \, \de s + \int_0^t e^{\bfA_N(t-s)} \, \de W_s^N. 
\end{align}
Discretizing in time, we arrive at the following recurrence formula. Let $Y_k^{N} = X_{k\Delta t} ^{N}$; then we have
\begin{align}
	Y_{k+1}^{N} = e^{\bfA_N \Delta t} Y_k^{N} + \bfA_N^{-1} \left( e^{\bfA_N\Delta t}-\Id\right) F_N(Y_k^{N}) + \int_{t_k}^{t_{k+1}} e^{\bfA_N (t_{k+1}-s)} \, \de W_s.
\end{align}
This recurrence equation can be decoupled into $N$ equations describing the evolutions of the coefficients of the solution. Let $Y_{k+1,i}^{N}$ and $F_N^i$ denote the $i$th components of $Y_{k+1}^{N}$ and $F_N$, respectively. Each component of $Y_{k+1}^{N}$ evolves according to
\begin{align}
 	Y_{k+1,i}^{N} = e^{-\lambda_i \Delta t}Y_{k,i}^{N} +\frac{1-e^{-\lambda_i \Delta t}}{\lambda_i} F_N^i(Y_k^{N}) + \sqrt{q_k}\int_{t_k}^{t_{k+1}} e^{-\lambda_i (t_{k+1}-s)} \, \de \beta_i(s).
 \end{align} 
As for the last integral, note that any stochastic integral where the integrand is not dependent on the Brownian motion is Gaussian \cite{oksendal2003stochastic}. Furthermore, any It\^o integral has mean zero (by virtue of being a martingale) and its variance can be computed via the It\^o isometry. That is, 
\begin{align}
	\Ex\left[ \left(\int_{t_{k}}^{t_{k+1}} e^{-\lambda_i (t_{k+1}-s)} \, \de \beta_i(s)\right)^2 \right] &= \int_{t_k}^{t_{k+1}} e^{-2\lambda_i(s-t_{k+1})} \de s \\
	& = \frac{1}{2\lambda_i}\left( 1-e^{-2\lambda_i \Delta t} \right).
\end{align}
Therefore, the numerical algorithm for simulating the SPDE is
\begin{align}
	Y_{k+1,i}^{N} = e^{-\lambda_i \Delta t}Y_{k,i}^{N} +\frac{1-e^{-\lambda_i \Delta t}}{\lambda_i} F_N^i(Y_k^{N}) +\sqrt{\frac{q_k}{2\lambda_i}(1-e^{-2\lambda_i \Delta t})} \Delta W_k^i \, ,
\end{align}
where $\Delta W_k^i \sim \mathcal{N}(0,1)$.
When simulating the stochastic advection-diffusion equation, in particular, we have $\bfA v = \alpha v_{xx}$ and $f(v) = bv_x $. 

\bibliographystyle{unsrt}
\bibliography{bibliotheque.bib}

\end{document}